\documentclass[a4paper,11pt]{article}
\pdfoutput=1 

\usepackage{jheppub} 

\usepackage[T1]{fontenc} 
\usepackage{comment}

\newcommand{\bi}{\begin{itemize}}
    \newcommand{\ei}{\end{itemize}}
\newcommand{\bea}{\begin{eqnarray}}
    \newcommand{\eea}{\end{eqnarray}}
\newcommand{\bt}{\begin{tabular}}
    \newcommand{\et}{\end{tabular}}
\newcommand{\bc}{\begin{center}}
    \newcommand{\ec}{\end{center}}

\newcommand{\be}{\begin{equation}}
    \newcommand{\ee}{\end{equation}}
\newcommand{\ba}{\begin{array}}
    \newcommand{\ea}{\end{array}}
\newcommand{\p}[1]{(\ref{#1})}
\newcommand{\lb}[1]{\label{#1}}

\def\bbox{{\,\lower0.9pt\vbox{\hrule \hbox{\vrule height 0.2 cm
                \hskip 0.2 cm \vrule height 0.2 cm}\hrule}\,}}
\newcommand{\dsl}{\pa \kern-0.5em /}

\newcommand{\nn}{\nonumber \\}

\makeatletter \@addtoreset{equation}{section} \makeatother

\def\slashchar#1{\setbox0=\hbox{$#1$}           
    \dimen0=\wd0                                 
    \setbox1=\hbox{/} \dimen1=\wd1               
    \ifdim\dimen0>\dimen1                        
    \rlap{\hbox to \dimen0{\hfil/\hfil}}      
    #1                                        
    \else                                        
    \rlap{\hbox to \dimen1{\hfil$#1$\hfil}}   
    /                                         
    \fi}

\pdfoutput=1

\title{{\boldmath Off-shell cubic hypermultiplet couplings to}\\
    $\mathcal{N}=2$ {\boldmath higher spin gauge superfields}}

\author[a,b,c]{Ioseph Buchbinder}
\author[d,e]{Evgeny~Ivanov,}
\author[d,e]{Nikita~Zaigraev}

\affiliation[a]{Center of Theoretical Physics, Tomsk State Pedagogical University,\\ 634061, Tomsk,  Russia}

\affiliation[b]{National Research Tomsk State University, \\634050, Tomsk, Russia}
\affiliation[c]{Lab of Theor. Cosmology, International Centre of Gravity
    and Cosmos,\\
    Tomsk State University of Control Systems and Radioelectronics,\\
    (TUSUR), 634050, Tomsk, Russia}
\affiliation[d]{Bogoliubov Laboratory of Theoretical Physics, JINR,\\141980 Dubna, Moscow region, Russia}
\affiliation[e]{Moscow Institute of Physics and Technology,\\ 141700 Dolgoprudny, Moscow region, Russia}

\emailAdd{joseph@tspu.edu.ru}
\emailAdd{ eivanov@theor.jinr.ru}
\emailAdd{nikita.zaigraev@phystech.edu}

\abstract{We construct manifestly  $4D, \mathcal{N}=2$ supersymmetric
    and gauge invariant  off-shell cubic couplings of matter hypermultiplets  to the higher integer spin gauge $\mathcal{N}=2$ multiplets
    introduced in {\tt \href{https://arxiv.org/abs/2109.07639}{arXiv:2109.07639 [hep-th]}}.
    The hypermultiplet is described by an analytic harmonic $4D, \mathcal{N}=2$
    superfield $q^{+}$ with the physical component spins
    ${\bf s} = (\frac{1}{2}\,, \;0)$ and an infinite number of auxiliary fields.
    The cubic coupling constructed has the schematic structure $q^+ \hat{{\cal H}}^{++}_{(s)} q^+$, where
    $\hat{{\cal H}}^{++}_{(s)}$ is a differential analytic operator of the highest degree $({\bf s} - 1)$ accommodating
    the massless gauge $\mathcal{N}=2$ multiplet with the highest spin  ${\bf s}$. For odd ${\bf s}$ the gauge group generators
    and couplings are proportional to ${\rm U}(1)_{PG}$ generator of the internal ${\rm SU}(2)_{PG}$ symmetry of the hypermultiplet and so
    do not exist  if ${\rm SU}(2)_{PG}$ is unbroken. If this ${\rm U}(1)_{PG}$ is
    identified with the central charge of $ 4D, \mathcal{N}=2$ supersymmetry, a mass for the
    hypermultiplet is generated and the odd ${\bf s}$ couplings vanish in the proper massless limit.
    For even ${\bf s}$ the higher-spin gauge transformations and cubic superfield couplings can be defined for both massive and massless (central-charge neutral) hypermultiplets
    without including ${\rm U}(1)_{PG}$ generator.
    All these features directly extend to the case of $n$ hypermultiplets with the maximal internal symmetry ${\rm USp}(2n) \times {\rm SU}(2)$.
}

\begin{document}
\maketitle

\flushbottom

\section{Introduction}
\label{sec:intro} The theory of higher spin fields
attracts vast attention due to its intimate relationships with both
quantum field theory and (super)string theory.
In particular, it was conjectured \cite{VAS} that the
superstring theory describes the ``massive'' phase of the massless higher
spin theory corresponding to the spontaneous breakdown of the
infinite-rank gauge symmetries of the latter, with masses of the
higher-spin excitations being generated  through some generalized
Higgs-type mechanism. Though the precise role of the higher-spin
fields in fundamental interactions is still unclear, it is expected
that the relevant  particles could dominate the string interactions
at high energies (see, e.g., \cite{Sagnotti:2011jdy}). Like
in other instances, the study of supersymmetric versions of various
higher-spin models is of pivotal importance, as supersymmetry can
shed more light on the hidden structures and properties of the
theory.
Since the higher-spin theory, like the superstring theory, predicts
infinite towers of fermions and bosons related by supersymmetry,
there arises the natural problem of describing these sets in the
appropriate field-theory language, where fields are combined into
supermultiplets and further into superfields. Thus, the issue of
describing supersymmetric theories of higher spins in the
superfield approach acquires high actuality.

The superfield formulations have a great privilege to provide an
explicit off-shell realization of supersymmetry.  They are  most
preferable for constructing invariant actions since the
transformations of given supersymmetry do  not depend on the choice
of a specific model. In four dimensions, there were developed  the
exhausting superfield formulations of $\mathcal{N}=1$ theories in
general and chiral $4D, \mathcal{N}=1 $ superspaces (see,
e.g., \cite{GGRS, BK}), as well as  of $\mathcal{N}=2$ theories in
the harmonic superspace and its analytic subspace with half the
original Grassmann coordinates \cite{18, HSS}.

In a recent  paper \cite{Buchbinder:2021ite} we have constructed the
completely off-shell manifestly ${\cal N}=2$ supersymmetric
superfield extension of arbitrary $4D$ integer-spin free massless
theory. The construction was based on the harmonic superspace method
\cite{18, HSS}, which so confirmed its universal role as the most
adequate and convenient approach to $4D, {\cal N}=2$ supersymmetric
field theories. The  theory constructed yields just  $\mathcal{N}=2$
off-shell supersymmetrization of the free massless
higher-spin field theory pioneered by Fronsdal \cite{FronsdalInteg}
and Fang and Fronsdal \cite{FronsdalHalfint}.

The natural next step is to set up interacting extensions of the
free theory. The construction of interactions in the theory of
higher spins is a very important, but highly difficult task. The only fully nonlinear constructions known at the moment are
limited to the Vasiliev's theory in (A)dS$_d$ \cite{Vasiliev:1990en,
Vasiliev:1992av, Vasiliev:2003ev} (for a review see \cite{V1, V2, V3,
Bekaert:2004qos, Didenko:2014dwa}) and the $3D$ Chern-Simons higher
spin gravity formulations \cite{Blencowe:1988gj,
Gaberdiel:2012uj, Ammon:2012wc, Gutperle:2011kf} (for a review see
\cite{Kiran:2014dfa}).
As for the interacting higher spin theories in the Lagrangian
approach, one of the possible obstructions here is provided by no-go
theorems imposing significant restrictions on the corresponding
interactions (for a review and references see, e.g.,
\cite{Bekaert:2010hw}). The statement of the no-go theorems relevant
to our consideration is that the constraints on the low-energy
scattering in flat space-time seemingly forbid massless particles
with spins $s>2$ to participate in interactions. However, in many
cases no-go theorems may be evaded  by relaxing various original
assertions.

The simplest higher spin interaction is described by a cubic vertex,
e.g., bilinear in the matter fields and of the first order
in gauge fields. At present, there is an extensive literature
related to the construction of cubic higher spin interactions. The
cubic  $s-s^\prime-s^{\prime\prime}$ couplings for
massless fields of arbitrary helicity in flat space-time have been
pioneered in \cite{Bengtsson:1983pd, Bengtsson:1983pg} in the
light-cone formalism\footnote{In \cite{Bengtsson:1983pg} the
corresponding cubic vertex was found for supermultiplets using a
light-cone superspace.}.
Generalizations of these results
are presented in \cite{Berends:1984rq, Berends:1985xx,
Bengtsson:1986kh, Fradkin:1991iy, Fradkin:1995xy, Metsaev:1993ap,
Metsaev:1993gx, Metsaev:1993mj, Metsaev:2007rn, Metsaev:2017cuz,
Metsaev:2018moa}. Higher-spin interactions with matter fields and
the corresponding conserved currents were firstly considered in
\cite{Berends:1984rq, Berends:1985xx} as the manifestly Poincar\'e
invariant vertices in the Lagrangian approach. Some more recent
results concerning higher spin cubic interaction can be
found in \cite{Buchbinder:2006eq, Fotopoulos:2007yq,
Fotopoulos:2007nm, Bekaert:2009ud, Manvelyan:2009vy,
Manvelyan:2010je, Manvelyan:2010jr, Manvelyan:2010wp,
Metsaev:2012uy, BL, BR, Fotopoulos:2008ka}.

The study of supersymmetric higher spin theories has been pioneered
in \cite{Courtright, Vasiliev} in the component approach
(for further development see, e.g., \cite{BKoutr, Z1, Z2, BS}
and the references therein). Recently, various methods have
been developed for deriving and exploring the supersymmetric cubic
vertices in on-shell supersymmetric formulations \cite{Metsaev1,
Metsaev2, Zinoiviev, BKTW}.

The superfield off-shell formulations of the $\mathcal{N}=1$
supersymmetric free massless higher spin theories in $4D$ flat and
AdS spaces were constructed for the first time in \cite{Kuz1, Kuz2,
Kuz3}. The superfield Lagrangian description of the free massless
${\cal N}=2$ higher spin supermultiplets in terms of ${\cal N}=1$
superfields was developed in \cite{KSG1,KSG2}. Superconformal free
higher spin field models in the ${\cal N}=2$ superfield approach
were discussed in \cite{Kuzenko:2017ujh, Kuzenko:2020,
Kuzenko:2021}. A superfield formulation of $\mathcal{N}=1$ free
massive higher spin fields was constructed in \cite{K}.
Supersymmetric $\mathcal{N}=1$ generalizations of the purely bosonic
cubic vertices with matter and the corresponding supercurrents were
explored in terms of $\mathcal{N}=1$ superfields in
\cite{Buchbinder:2017nuc, HK1, HK2, Buchbinder:2018gle, BHK,
Buchbinder:2018wwg, Buchbinder:2018wzq, Buchbinder:2019yhl}.

Let us turn back to $4D, \mathcal{N}=2$ supersymmetric theory.
The most general lower spin $4D, \mathcal{N}=2$ matter
multiplet is hypermultiplet with the physical component spins $s = (\frac12, \,0)$.  The off-shell hypermultiplet requires
an infinite number of auxiliary fields and admits an unconstrained
formulation only in $\mathcal{N}=2$ harmonic superspace
\cite{18}. So if one intends to study off-shell
$\mathcal{N}=2$ cubic vertices, it is unavoidable to use the
harmonic superspace approach.

In this paper we construct, for the first time, the off-shell
manifestly $\mathcal{N}=2$ supersymmetric cubic couplings
$\mathbf{\frac{1}{2}-\frac{1}{2} - s}$  of an arbitrary integer higher  spin $\mathbf{s}$ gauge $\mathcal{N}=2$ multiplet to the
hypermultiplet matter in $4D, \mathcal{N}=2$ harmonic
superspace\footnote{We use the bold letters $\mathbf{s}$ to denote
the highest spin (``superspin'') in $\mathcal{N}=2$
supermultiplet.}. The couplings are linear in the analytic
gauge superfields, bilinear in the analytic hypermultiplet $q^+$
superfields and are written as integrals over the analytic harmonic
superspace. An interesting new peculiarity is the necessity of
breaking the Pauli-G\"ursey ${\rm SU}(2)_{PG}$ symmetry of the free
hypermultiplet action down to ${\rm U}(1)$ for odd spins ${\bf s}$ and the explicit presence
of this  ${\rm U}(1)$ generator in the gauge transformations and the
relevant superfield cubic actions. If this
${\rm U}(1)$ generator is identified with the central charge of
rigid $4D, {\cal N}=2$ supersymmetry, it induces a mass for the
hypermultiplet and only even spin couplings survive in the massless limit.
As regards the higher-spin gauge superfields, we deal with the same linearized
actions for them as in ref. \cite{Buchbinder:2021ite}.

The cubic couplings of the hypermultiplet to $\mathcal{N}=2$ higher
spin gauge superfield constructed in the present paper are based on
the gauge principle: the interactions are resulted from localization
of global symmetry. We begin with the free hypermultiplet theory in
harmonic superspace and find for this theory an infinite set of
higher-derivative global symmetry transformations. The localization
of these transformations in the harmonic analytic superspace
requires introducing the corresponding higher spin gauge harmonic
superfields with the gauge group already described in
\cite{Buchbinder:2021ite}. As a result, we arrive at the manifestly
$\mathcal{N}=2$ supersymmetric cubic interactions of the
hypermultiplet with the higher spin gauge superfields, in the first
order in the latter\footnote{A cubic coupling of massless on-shell
hypermultiplet to higher spin superconformal gauge superfields
 was recently studied in \cite{Kuzenko:2021}. Here we deal entirely with off-shell interactions
and do not concern superconformal theories at all.}.

The paper is organized as follows. Section  \ref{sec:hs} is devoted
to a brief description of a completely off-shell invariant action
for the free massless ${\cal N}=2$ gauge theory with an arbitrary
maximal integer spin ${\bf s}$ of the supermultiplet. In
section \ref{sec:22} we introduce the analytic
$\mathcal{N}=2$ invariant differential superfield operators
associated with the higher-derivative hypermultiplet gauge
transformations. Just these operators are the basic ingredients  of our
construction. In section \ref{sec:hyper} we present the detailed
derivation of the manifest form of $\mathbf{\frac{1}{2}-\frac{1}{2}
- s}$ couplings of $\mathcal{N}=2$ higher spins to the
hypermultiplet for $\mathbf{s} = 1, 2, 3, 4$, as well as the
corresponding global and local transformations of the
hypermultiplet.  A generalization to arbitrary even and odd ${\cal
N}=2$ spins is the subject  of section \ref{sec:hyperGen}. In
section \ref{sec:summary} we summarize the results, briefly describe
a generalization to an arbitrary number of hypermultiplets (with the internal symmetry ${\rm USp}(2n)$) and discuss
possible directions of further development of the material presented.

\section{$\mathcal{N}=2$ supersymmetric higher spins in harmonic superspace}
\label{sec:hs}

We will deal with ${\cal N}=2$ harmonic superspace (HSS) in the analytic
basis parametrized by the following set of coordinates \cite{HSS,18}
\be
Z =
\big(x^m, \theta^{+\mu}, \bar\theta^{+\dot\mu}, u^\pm_i,
\theta^{-\mu}, \bar\theta^{-\dot\mu}\big) \equiv \big(\zeta,
\theta^{-\mu}, \bar\theta^{-\dot\mu}\big), \lb{HSS}
\ee
\be
\zeta = \big(x^m,
\theta^{+\mu}, \bar\theta^{+\dot\mu}, u^\pm_i\big)\,, \lb{AHSS}
\ee
where the
standard notation of ref. \cite{18} is used. In particular,
$u^\pm_i$ are harmonic variables parametrizing the internal sphere
$S^2$, $u^{+i}u^-_i =1$, the indices $\pm$ denote the harmonic
$U(1)$ charges of various quantities and the index $i = 1, 2$ is the
doublet index of the automorphism group ${\rm SU}(2)_{aut}$  acting only on the
harmonic variables. The set \p{HSS} is closed under the rigid ${\cal N}=2$
supersymmetry transformations
\be
\delta_\epsilon x^{\alpha\dot\alpha} = -2i
\big(\epsilon^{- \alpha}\bar\theta^{+ \dot\alpha} + \theta^{+ \alpha}\bar\epsilon^{- \dot\alpha}\big), \quad \delta_\epsilon\theta^{\pm \hat \mu} =
\epsilon^{\pm \hat \mu}\,, \quad \delta_\epsilon u^{\pm}_i = 0\,, \quad
\epsilon^{\pm \hat \mu} = \epsilon^{\hat \mu i } u^\pm_i\,,
\lb{N2SUSY}
\ee
where we employed the condense notation\footnote{Hereafter, we use the notations
    $\hat{\mu} \equiv (\mu, \dot{\mu})$, $\partial^\pm_{\hat{\mu}} =
    \partial/ \partial \theta^{\mp \hat{\mu}}$, $(\theta^{\hat{+}})^2
    \equiv (\theta^+)^2 - (\bar{\theta}^+)^2$ and
    $\partial_{\alpha\dot\alpha} =
    \sigma^m_{\alpha\dot\alpha}\partial_m$. The summation rules are
    $\psi\chi = \psi^\alpha \chi_\alpha, \bar\psi\bar\chi =
    \bar\psi_{\dot\alpha} \bar\chi^{\dot\alpha}$, Minkowski metric is
    ${\rm diag} (1, -1, -1, -1)$ and $\Box = \partial^m\partial_m =
    \frac12 \partial^{\alpha\dot\alpha}\partial_{\alpha\dot\alpha}$.}, $\hat\mu =
(\mu, \dot\mu)$. These transformations also leave intact the
harmonic analytic subspace $\zeta$ \p{AHSS}.

Both the harmonic superspace and its analytic subspace are
self-conjugated under the generalized tilde involution \cite{18}:
\begin{equation}\label{tilde}
    \widetilde{x^m} = x^m\,, \qquad \widetilde{\theta^\pm_\alpha} = \bar{\theta}^\pm_{\dot{\alpha}} \,, \qquad
    \widetilde{\bar{\theta}^\pm_{\dot{\alpha}}} = - \theta^\pm_\alpha \,, \qquad
    \widetilde{u^{\pm i}} = - u_i^\pm\,, \qquad
    \widetilde{u^\pm_i} = u^{\pm i}\,.
\end{equation}

The HSS formulation of ${\cal N}=2$ gauge theories coupled to the hypermultiplet
matter uses an extension
of the HSS \p{HSS} by a fifth coordinate $x^5$,
\be
Z \Longrightarrow (Z, x^5)\,\supset \, (\zeta, x^5)\,, \qquad \widetilde{x^5} = x^5\,, \lb{Ext5}
\ee
with the following
analyticity-preserving transformation law under ${\cal N}=2$
supersymmetry,
\be
\delta_\epsilon x^5 = 2i\big(\epsilon^-\theta^+ -
\bar\epsilon^-\bar\theta^+ \big). \lb{Tranfifth}
\ee
This coordinate
can be interpreted as associated with the central charge in ${\cal N}=2$ Poincar\'e superalgebra.

An important ingredient of the HSS formalism is the harmonic
derivatives $\mathcal{D}^{++}$ and $\mathcal{D}^{--}$ which have the following form in
the analytic basis
\begin{eqnarray}
    && \mathcal{D}^{++} = \partial^{++} - 2i \theta^{+\rho} \bar{\theta}^{+\dot{\rho}} \partial_{\rho\dot{\rho}} + \theta^{+\hat{\mu}} \partial^{+}_{\hat{\mu}}
    +
    i (\theta^{\hat{+}})^2 \partial_5\,,  \lb{Dflat+} \\
    && \mathcal{D}^{--} = \partial^{--}- 2i \theta^{-\rho} \bar{\theta}^{-\dot{\rho}} \partial_{\rho\dot{\rho}} + \theta^{-\hat{\mu}} \partial^{-}_{\hat{\mu}}
    +
    i (\theta^{\hat{-}})^2 \partial_5\,, \lb{Dflat} \\
    && [\mathcal{D}^{++}, \mathcal{D}^{--}] = D^0\,, \quad D^0 = \partial^0 + \theta^{+ \hat\mu}\partial^-_{\hat\mu}
    -\theta^{- \hat\mu}\partial^+_{\hat\mu}\,.\lb{Flatness}
\end{eqnarray}
Here we used the standard notations for the partial derivatives with respect to  harmonic variables:
\begin{equation}
    \partial^{++} = u^{+i} \frac{\partial}{\partial u^{-i}}\,,\qquad
    \partial^{--} = u^{-i} \frac{\partial}{\partial u^{+i}}\,,\qquad
    \partial^0 = u^{+ i}\frac{\partial}{\partial u^{+i}} - u^{- i}\frac{\partial}{\partial u^{-i}}\,.
\end{equation}

The crucial difference between the derivatives $\mathcal{D}^{++}$ and $\mathcal{D}^{--}$ is that $\mathcal{D}^{++}$ preserves analyticity,
while $\mathcal{D}^{--}$ does not. All superfields except for the hypermultiplet are assumed to be $x^5$-independent, while the action of $\partial_5$
on the hypermultiplet is identified with the generator of some ${\rm U}(1)$ isometry of the free hypermultiplet action (see Section \ref{sec:hyper}).

For what follows it will be useful to present the ``passive'' transformations of various partial derivatives under ${\cal N}=2$ supersymmetry:
\begin{eqnarray}
&& \delta_\epsilon \partial^-_\alpha = 2i \bar\epsilon^{-\dot\alpha}\partial_{\alpha\dot\alpha} - 2i \epsilon^-_\alpha \partial_5\,, \quad
\delta_\epsilon \partial^-_{\dot\alpha} = -2i \epsilon^{-\alpha}\partial_{\alpha\dot\alpha}
- 2i \bar\epsilon^-_{\dot\alpha} \partial_5\,,\nonumber \\
&& \delta_\epsilon \partial_{\alpha\dot\alpha} = \delta_\epsilon \partial_{5} = 0\,. \lb{TranPart}
\end{eqnarray}

\subsection{Analytic prepotentials and invariant action}

In this section we  summarize the main results of \cite{Buchbinder:2021ite}. We present all the necessary information
about description of $\mathcal{N}=2$ higher spin supermultiplets in harmonic superspace. Detailed examples of spin ${\bf 2}$ and spin ${\bf 3}$ description
were given in \cite{Buchbinder:2021ite}, here we will at once deal with the general structure of arbitrary spin $\mathbf{s}$ theory.

The $\mathcal{N}=2$ gauge multiplet  of highest integer spin $\mathbf{s}$ ($\mathbf{s}\geq 2$) \footnote{One can include the spin ${\bf s =1}$
    into this hierarchy as well:  it is described by a single analytic
    superfield $h^{++}$ and encompasses the Abelian gauge ${\cal N}=2$
    multiplet (spins $({ 1, 1/2, 1/2, 0})$ on shell).} is accommodated by two real analytic bosonic superfields
\begin{subequations}\label{eq:q}
\begin{equation}\label{eq:q:1}
    h^{++}_{\alpha(s-1)\dot{\alpha}(s-1)} (\zeta), \;\;\;\;\;   h^{++ 5}_{\alpha(s-2)\dot{\alpha}(s-2)}(\zeta)
\end{equation}
and two conjugated analytic spinor superfields
\begin{equation}
    \label{eq:q:2}
    h^{+++}_{\alpha(s-1)\dot{\alpha}(s-2)}(\zeta), \;\;\;\;\;   h^{+++}_{\alpha(s-2)\dot{\alpha}(s-1)}(\zeta),
\end{equation}
\end{subequations}
where the symbols $\alpha(s)$ and $\dot{\alpha}(s)$ stand for totally symmetric combinations of s spinor indices,
$\alpha(s): = (\alpha_1, \dots \alpha_s)$, $\dot{\alpha}(s): = (\dot{\alpha}_1, \dots \dot{\alpha}_s)$. The index $5$ in \eqref{eq:q:1} is inherited from the spin ${\bf 2}$
case where $h^{++ 5}$ is the linearized form of the analytic vielbein of ${\cal N}=2$ supergravity associated with the central
charge $\partial_5$. The corresponding gauge group
is implemented  by the transformations \footnote{Our normalization of $h^{++\alpha(s-1)\dot\alpha(s-1)}$ and $\lambda^{\alpha(s-1)\dot\alpha(s-1)}$ differs by the factor 2 from
the one used in \cite{Buchbinder:2021ite}.}:
\begin{equation}\label{Gauge_s}
    \begin{split}
         \delta_\lambda h^{++\alpha(s-1)\dot\alpha(s-1)} =\,& \mathcal{D}^{++} \lambda^{\alpha(s-1)\dot\alpha(s-1)} \\&+
        2i \big[\lambda^{+\alpha(s-1)(\dot\alpha(s-2)}\bar\theta^{+\dot\alpha_{s-1})} + \theta^{+(\alpha_{s-1}} \bar\lambda^{+\alpha(s-2))\dot\alpha(s-1)} \big], \\
        \delta_\lambda h^{++ 5\,\alpha(s-2)\dot\alpha(s-2)} =\,& \mathcal{D}^{++} \lambda^{\alpha(s-2)\dot\alpha(s-2)} \\&-
        2i\,\big[\lambda^{+(\alpha(s-2)\alpha_{s-1})\dot\alpha(s-2)} \theta^+_{\alpha_{s-1}} +
        \bar\lambda^{+(\dot\alpha(s-2)\dot\alpha_{s-1})\alpha(s-2)} \bar\theta^+_{\dot\alpha_{s-1}} \big], \\
         \delta_\lambda  h^{++\alpha(s-1)\dot\alpha(s-2)+} =\,& \mathcal{D}^{++}\lambda^{+\alpha(s-1)\dot\alpha(s-2)}\,, \\
         \delta_\lambda h^{++\dot\alpha(s-1)\alpha(s-2)+} =\,&
        \mathcal{D}^{++}\bar\lambda^{+\dot\alpha(s-1)\alpha(s-2)}\,.
    \end{split}
\end{equation}
These transformations can be used to choose the appropriate WZ gauge and then show that the physical multiplet involves spins $( s, s-\frac{1}{2}, s-\frac{1}{2}, s-1)$:
\begin{equation}\label{WZ3}
    \begin{split}
     &h^{++\alpha(s-1)\dot{\alpha}(s-1)}
    =
    -2i \theta^{+\rho} \bar{\theta}^{+\dot{\rho}} \Phi^{\alpha(s-1)\dot{\alpha}(s-1)}_{\rho\dot{\rho}}
    +  (\bar{\theta}^+)^2 \theta^+ \psi^{\alpha(s-1)\dot{\alpha}(s-1)i}u^-_i  \\
    & \;\; \;\;\;\; \;\;\;\; \;\;\;\; \;\;\; \;\;\;\; \;\;\;\;\;+\, (\theta^+)^2 \bar{\theta}^+ \bar{\psi}^{\alpha(s-1)\dot{\alpha}(s-1)i}u_i^-
    +  (\theta^+)^2 (\bar{\theta}^+)^2 V^{\alpha(s-1)\dot{\alpha}(s-1)(ij)}u^-_iu^-_j\,, \\
    &  h^{++ 5\,\alpha(s-2)\dot{\alpha}(s-2)} =
    -2i \theta^{+\rho} \bar{\theta}^{+\dot{\rho}} C^{\alpha(s-2)\dot{\alpha}(s-2)}_{\rho\dot{\rho}}
    + (\bar{\theta}^+)^2 \theta^+ \rho^{\alpha(s-2)\dot{\alpha}(s-2)i}u^-_i
    \\
    & \;\; \;\;\;\; \;\;\;\; \;\;\;\; \;\;\; \;\;\;\; \;\;\;\;\;+\,
    (\theta^+)^2 \bar{\theta}^{+} \bar{\rho}^{\alpha(s-2)\dot{\alpha}(s-2)i}u_i^-
    + (\theta^+)^2 (\bar{\theta}^+)^2 S^{\alpha(s-2)\dot{\alpha}(s-2)(ij)}u^-_iu^-_j\,, \\
    &  h^{++\alpha(s-1)\dot{\alpha}(s-2)+} = (\theta^+)^2 \bar{\theta}^+_{\dot{\mu}} P^{\alpha(s-1)\dot{\alpha}(s-2)\dot{\mu}}
    \\
    & \;\; \;\;\;\; \;\;\;\; \;\;\;\; \;\;\; \;\;\;\; \;\;\;\;\;
    +  \left(\bar{\theta}^+\right)^2 \theta^+_\nu \left[\varepsilon^{\nu(\alpha} M^{\alpha(s-2))\dot{\alpha}(s-2)} + T^{\dot{\alpha}(s-2)(\alpha(s-1)\nu)}\right]
    \\
    & \;\; \;\;\;\; \;\;\;\; \;\;\;\; \;\;\; \;\;\;\; \;\;\;\;\;+\,
    (\theta^+)^2 (\bar{\theta}^+)^2 \chi^{\alpha(s-1)\dot{\alpha}(s-2)i}u^-_i\,, \\
    &  h^{++\dot{\alpha}(s-1)\alpha(s-2)+} = \widetilde{\left(h^{++\alpha(s-1)\dot{\alpha}(s-2)+}\right)}\,.
    \end{split}
\end{equation}
Here the fields
\begin{equation}\label{physical fields}
    \Phi^{\alpha(s-1)\dot{\alpha}(s-1)}_{\rho\dot{\rho}},\;\;\;\;\;
    \psi^{\alpha(s-1)\dot{\alpha}(s-1)i}, \;\;\;\;\;
    \bar{\psi}^{\alpha(s-1)\dot{\alpha}(s-1)i}, \;\;\;\;\;
    C^{\alpha(s-2)\dot{\alpha}(s-2)}_{\rho\dot{\rho}}
\end{equation}
are physical and describe the bosonic spin $s$, a doublet of the fermionic $s-\frac{1}{2}$ spin fields and the bosonic spin $s-1$. All other fields are auxiliary.

For discussion of the residual gauge transformations and further features of the component reduction we refer
the reader to \cite{Buchbinder:2021ite}, where all details in the ${\bf s=2}$ and ${\bf s=3}$ cases were presented. It was shown there
that the residual gauge transformations completely coincide with the corresponding transformations for the physical
 bosonic \cite{FronsdalInteg} and fermionic fields \cite{FronsdalHalfint}.

To construct the invariant linearized actions, one must introduce the negatively charged potentials
\begin{equation}
    \begin{split}
    h^{--}_{\alpha(s-1)\dot\alpha(s-1)}(Z),& \;\;\;\;\;
    h^{-- 5}_{ \alpha(s-2)\dot\alpha(s-2)}(Z)\,, \;
    \\
    h^{-- +}_{\alpha(s-1)\dot\alpha(s-2)}(Z),& \;\;\;\;\;
    h^{--+}_{\dot\alpha(s-1)\alpha(s-2)}(Z).
    \end{split} \lb{Sets2}
\end{equation}
They are defined as solutions of the harmonic flatness conditions:
\begin{equation}
    \begin{cases}
        \mathcal{D}^{++} h^{-- \alpha(s-1)\dot\alpha(s-1)}-
        \mathcal{D}^{--} h^{++\alpha(s-1)\dot\alpha(s-1)} 
        \\\qquad\qquad\qquad\qquad+
        2i \left(h^{--\alpha(s-1)(\dot{\alpha}(s-2)+}\bar{\theta}^{+\dot{\alpha})} + \theta^{+(\alpha} h^{--\alpha(s-2))\dot{\alpha}(s-1)+} \right)= 0\,,
        \\
        \mathcal{D}^{++} h^{-- 5\,\alpha(s-2)\dot\alpha(s-2)} -
        \mathcal{D}^{--} h^{++ 5\,\alpha(s-2)\dot\alpha(s-2)} \\\qquad\qquad\qquad\qquad-
        2i \left(h^{--\alpha(s-1)\dot{\alpha}(s-2)+}\theta^+_\alpha + \bar{\theta}^{+}_{\dot{\alpha}} h^{--\alpha(s-2)\dot{\alpha}(s-1)+} \right) = 0\,,
        \\
        \mathcal{D}^{++}h^{--\alpha(s-1)\dot\alpha(s-2)+} -
        \mathcal{D}^{--}h^{++\alpha(s-1)\dot\alpha(s-2)+} = 0\,,
        \\
        \mathcal{D}^{++}h^{--\dot\alpha(s-1)\alpha(s-2)+} -
        \mathcal{D}^{--}h^{++\dot\alpha(s-1)\alpha(s-2)+} = 0 \,.
    \end{cases}
\end{equation}
These extra non-analytic potentials will not be present in the Lagrangian of cubic interactions with
hypermultiplet and are needed only for constructing the invariant actions of the gauge superfields.

The bosonic analytic prepotentials \eqref{eq:q:1}, \eqref{eq:q:2} and the negatively charged potentials \eqref{Sets2}
have non-standard transformation laws under $\mathcal{N}=2$ global supersymmetry
\footnote{$\delta_\epsilon$ is the passive transformation, which differs from the active transformation $\delta_\epsilon^*$
by the ``transport term'', $\delta^*_\epsilon = \delta_\epsilon -\delta_\epsilon Z^M \partial_M$, where $M = (\alpha\dot{\alpha}, \hat{\mu}+, 5)$.
Though the gauge superfields do not depend on $x^5$, $\partial_5 h^{\pm\pm\dots}=0$, the derivative $\partial_5$ can be non-zero on the matter hypermultiplet superfields,
see section 3.}:
\begin{subequations}\label{susy}
\begin{equation}\label{susy1}
    \begin{split}
\delta_\epsilon h^{\pm\pm\alpha(s-1)\dot\alpha(s-1)} = -2i\big[h^{\pm\pm\alpha(s-1)(\dot\alpha(s-2)+}\bar\epsilon^{-\dot\alpha_{s-1})}-
h^{\pm\pm\dot\alpha(s-1)(\alpha(s-2)+}\,\epsilon^{-\alpha_{s-1})}
\big]\,, \\
\delta_\epsilon h^{\pm\pm 5\,\alpha(s-2)\dot\alpha(s-2)} =2i\big[h^{\pm\pm(\alpha(s-2)\alpha_{s-1})
    \dot\alpha(s-2)+}\epsilon^{-}_{\alpha_{s-1}} +
h^{\pm\pm\alpha(s-2)(\dot\alpha(s-2)\dot\alpha_{s-1})+}\,\bar\epsilon^{-}_{\dot{\alpha}_{s-1}}
\big] \,.
    \end{split}
\end{equation}
Spinor potentials have the standard $\mathcal{N}=2$ superfield passive transformation rules\footnote{In the full nonlinear case of
hypothetical ``higher-spin ${\cal N}=2$ supergravities'' the rigid ${\cal N}=2$ supersymmetry is expected to be included in the general gauge transformations.}:
\begin{equation}\label{susy2}
    \delta_\epsilon h^{\pm\pm\alpha(s-1)\dot\alpha(s-2)+} = 0,
    \;\;\;\;\;\;\;\;
    \delta_\epsilon h^{\pm\pm\dot\alpha(s-1)\alpha(s-2)+} = 0\,.
\end{equation}
\end{subequations}

As the building blocks of ${\cal N}=2$ supersymmetric actions, one defines non-analytic ${\cal N}=2$ singlet
superfields,
\begin{equation}
    \begin{split}
 G^{\pm\pm\alpha(s-1)\dot\alpha(s-1)} =\,&
h^{\pm\pm\alpha(s-1)\dot\alpha(s-1)} + 2i
\big[h^{\pm\pm\alpha(s-1)(\dot\alpha(s-2)+}\bar\theta^{-\dot\alpha_{s-1})}
\\&-
h^{\pm\pm\dot\alpha(s-1)(\alpha(s-2)+}\,\theta^{-\alpha_{s-1})}
\big], \\  G^{\pm\pm 5\,\alpha(s-2)\dot\alpha(s-2)} =\,&
h^{\pm\pm 5\,\alpha(s-2)\dot\alpha(s-2)} - 2i
\big[h^{\pm\pm(\alpha(s-2)\alpha_{s-1})
    \dot\alpha(s-2)+}\theta^{-}_{\alpha_{s-1}}
\\&+
h^{\pm\pm\alpha(s-2)(\dot\alpha(s-2)\dot\alpha_{s-1})+}\,\bar\theta^{-}_{\dot{\alpha}_{s-1}}
\big].
    \end{split}
 \lb{Ggen}
\end{equation}
It is straightforward to be convinced that indeed $\delta_{\epsilon}G^{\pm\pm\cdots} = 0$.

The newly introduced superfields satisfy the harmonic flatness conditions
\bea
&& \mathcal{D}^{++}G^{--\alpha(s-1)\dot\alpha(s-1)} - \mathcal{D}^{--}G^{++\alpha(s-1)\dot\alpha(s-1)} =0\,, \nn
&& \mathcal{D}^{++}G^{-- 5\,\alpha(s-2)\dot\alpha(s-2)} - \mathcal{D}^{--}G^{++ 5\,\alpha(s-2)\dot\alpha(s-2)} = 0 \nonumber
\eea
and transform under the gauge group as
\be
\delta_\lambda G^{\pm\pm\alpha(s-1)\dot\alpha(s-1)} =
D^{\pm\pm}\Lambda^{\alpha(s-1)\dot\alpha(s-1)}\,, \quad
\delta_\lambda G^{\pm\pm 5\,\alpha(s-2)\dot\alpha(s-2)} =
D^{\pm\pm}\Lambda^{\alpha(s-2)\dot\alpha(s-2)}\,,
\ee
where $\Lambda$'s are composed out of the analytic gauge parameters $\lambda$'s and the coordinates $\theta^{-\hat{\alpha}}$:
\bea
&&\Lambda^{\alpha(s-1)\dot\alpha(s-1)} =
\lambda^{\alpha(s-1)\dot\alpha(s-1)} +
2i\big[\lambda^{+\alpha(s-1)(\dot\alpha(s-2)}\bar\theta^{-\dot\alpha_{s-1})}
- \bar\lambda^{+\dot\alpha(s-1)(\alpha(s-2)} \theta^{-\alpha_{s-1})}
\big],\nn
&&\Lambda^{\alpha(s-2)\dot\alpha(s-2)} =
\lambda^{\alpha(s-2)\dot\alpha(s-2)}
-2i\big[\lambda^{+(\alpha(s-2)\alpha_{s-1})\dot\alpha(s-2)}\theta^{-}_{\alpha_{s-1}}
\nn
&&\;\;\;\;\;\;\;\;\;\;\;\;\;\;\;\;\;\;\;\;\;\;\;-\,
\bar\theta^{-}_{\dot\alpha_{s-1}}
\bar\lambda^{+(\dot\alpha(s-2)\dot\alpha_{s-1})\alpha(s-2)}\big]\,.
\lb{LambS}
\eea

The $\mathcal{N}=2$ gauge invariant action, up to a normalization factor, has the universal form
for any $s$:
\bea
&& S_{(s)} = (-1)^{s+1} \frac{1}{\kappa_s^2}  \int d^4x
d^8\theta du \,\Big\{G^{++
    \alpha(s-1)\dot\alpha(s-1)}G^{--}_{\alpha(s-1)\dot\alpha(s-1)} \nn
&&\;\;\;\;\;\;\;\;\;\;\;\;\;\;\;\;\;\;\;\;\;\;\;\quad \quad \quad\quad \quad \quad \quad   +\,
G^{++ 5\,\alpha(s-2)\dot\alpha(s-2)}G^{--5}_{\alpha(s-2)\dot\alpha(s-2)}
\Big\},  \lb{ActionsGen}
\eea
 where $\kappa_s$ is a higher spin analog of Newton's constant. The ${\cal N}=2$ supersymmetry of \p{ActionsGen} is
manifest, while the gauge invariance can be checked by bringing the gauge
variation of  $S_{(s)}$ to the form
\bea\label{higher spin action}
&& \delta_\lambda S_{(s)} = 2(-1)^{s+1} \frac{1}{\kappa_s^2}
\int d^4x d^8\theta du\;
\Big\{\mathcal{D}^{--}\Lambda^{\alpha(s-1)\dot\alpha(s-1)}G^{++
}_{\alpha(s-1)\dot\alpha(s-1)} \nn &&\;\;\;\;\;\;\; \quad \quad \quad \quad \quad \quad \quad \quad \quad \quad \quad \quad \quad +\,
\mathcal{D}^{--}\Lambda^{\alpha(s-2)\dot\alpha(s-2)}G^{++ 5}_{\alpha(s-2)\dot\alpha(s-2)}\Big\}
\eea
and proceeding further  by taking into account  that the $++$-potentials in \eqref{Ggen} are linear in the negatively charged coordinates  with the analytic coefficients.
The straightforward calculation immediately yields that $\delta_\lambda S_{(s)}=0$.

The dimension of the coupling constant $\kappa_s$ is fixed by the dimension of the analytic prepotentials. In mass units,
the ``engineering'' dimension of the analytic gauge potentials is $[h^{++\alpha(s-1)\dot{\alpha}(s-1)}] = [h^{++ 5\alpha(s-2)\dot{\alpha}(s-2)}] = - (s-1)$,
$[h^{++\alpha(s-1)\dot{\alpha}(s-2)+}] = -s + \frac32 $, whence $[\kappa_s] = -(s-1)$.
These dimensions differ from the canonical ones. For example, the engineering dimension of the physical spin $s$ field $\Phi^{\alpha(s-1)\dot{\alpha}(s-1)}_{\rho\dot{\rho}}$
in \eqref{physical fields} is $[\Phi^{\alpha(s-1)\dot{\alpha}(s-1)}_{\rho\dot{\rho}}] = -(s-2)$, while its canonical dimension is $+1$.
The canonical dimensions can be always restored by redefining the superfields in the set \eqref{eq:q:1} and their negatively charged counterparts
as $h^{\pm\pm\dots} = {\kappa_s} h^{\pm\pm\dots}_{\rm can}$.
In what follows, it will be more convenient to deal with the gauge superfields of engineering  dimensions.

Summarizing,  here we have repeated the main results of \cite{Buchbinder:2021ite}. We presented the transformation laws of the spin ${\bf s}$ superfields under both the gauge
group \eqref{Gauge_s} and $\mathcal{N}=2$ global supersymmetry \eqref{susy1}, \eqref{susy2} and then wrote down the invariant action \eqref{ActionsGen}
which is valid for any ${\bf s}$.

\subsection{Gauge transformations via differential operators}\label{sec:22}
In this section, we define analytic, $\mathcal{N}=2$ supersymmetry-preserving
differential operators which encode all the analytic prepotentials of the spin ${\bf s}$ higher spin supermultiplet.
We will start with the case of $\mathcal{N}=2$ Einstein supergravity (corresponding to ${\bf s}=2$) and then generalize this construction to
an arbitrary higher spin $\mathcal{N}=2$ supermultiplet.
These operators will play the key role in the construction of cubic couplings of hypermultiplet to higher spin gauge multiplets.

\subsubsection{Differential operators for gauge transformations in $\mathcal{N}=2$ Einstein \break supergravity}

One of the underlying  principles of Einstein $\mathcal{N}=2$
supergravity in harmonic superspace \cite{18}
is the covariantization of flat harmonic derivative \eqref{Dflat+}
\begin{eqnarray}
&&    \mathcal{D}^{++} \; \to \;\
    \mathfrak{D}^{++} = \mathcal{D}^{++} + \hat{\mathcal{H}}^{++}_{(2)}, \nonumber \\
&&    \hat{\mathcal{H}}^{++}_{(2)} := h^{++\alpha\dot{\alpha}} \partial_{\alpha\dot{\alpha}} + h^{++\hat{\mu}+} \partial^-_{\hat{\mu}} + h^{++5}\partial_5\,, \label{Hspin2}  \\
&&[\partial^+_{\hat{\mu}}, \hat{\mathcal{H}}^{++}_{(2)}] = 0\,.
\end{eqnarray}
Analytic superfields $h^{++\alpha\dot{\alpha}}, h^{++\hat{\mu}+}$ and $h^{++5}$ are unconstrained analytical prepotentials of $\mathcal{N}=2$ Einstein supergravity.
One can easily check that the operator $\hat{\mathcal{H}}^{++}_{(2)}$ is invariant under $\mathcal{N}=2$ global supersymmetry \eqref{TranPart} accompanied
by \eqref{susy1} and \eqref{susy2} specialized to the case ${\bf s}=2$. Note that the operator $\hat{\mathcal{H}}^{++}_{(2)}$ and its arbitrary spin $s$  generalizations
$\hat{\mathcal{H}}^{++}_{(s)}$ are dimensionless and the engineering dimensions of the component gauge superfields are uniquely
fixed by the dimensions of the related (higher-order) derivatives.

Fundamental gauge group of Einstein $\mathcal{N}=2$ supergravity
(its ``minimal'' version \cite{Fradkin:1979cw}) is the following  group of
superdiffeomorphisms of the analytic harmonic superspace:
\begin{eqnarray}
   && \delta_\lambda x^{\alpha\dot{\alpha}} = \lambda^{\alpha\dot{\alpha}} (\zeta),
    \;
    \delta_\lambda x^5 = \lambda^5(\zeta),
    \;
    \delta_\lambda \theta^{+\hat{\mu}} = \lambda^{+\hat{\mu}}(\zeta),\nonumber \\
&&\delta_\lambda u^{\pm i} = 0.
\end{eqnarray}

It will be useful for future application to realize these transformations on the gauge superfields as the ``active'' ones using the differential operator:
\begin{equation}\label{diffL}
    \hat{\Lambda}_{(2)} := \lambda^{\alpha\dot{\alpha}} \partial_{\alpha\dot{\alpha}}
    +
     \lambda^{+\hat{\mu}}\partial^-_{\hat{\mu}}
    +
    \lambda^5 \partial_5 := \Lambda^{M}\partial_M\,.
\end{equation}
Hereafter we use the notations $M = (\alpha\dot{\alpha}, \hat{\mu}+, 5)$.
By construction, the full covariant harmonic derivative is invariant under the action of gauge group of supergravity:
\begin{equation}\label{Dfull}
    \delta_\lambda \mathfrak{D}^{++} = 0\,.
\end{equation}
From \eqref{Dfull} one can easily deduce the active transformation laws for the analytic prepotentials \eqref{Gauge_s}
\begin{eqnarray}
    \delta^*_\lambda \mathfrak{D}^{++} = \delta^*_\lambda \hat{\mathcal{H}}^{++}_{(2)} = \delta^*_\lambda h^{++M} \partial_{M}= [\mathfrak{D}^{++}, \lambda^{M} \partial_{M}]\,. \label{Actives2}
\end{eqnarray}
For instance,
\bea
\delta^*_{\lambda} h^{++\alpha\dot\alpha} = \mathfrak{D}^{++}\lambda^{\alpha\dot\alpha} + 2i\big(\lambda^{+\alpha}\bar{\theta}^{+\dot\alpha} - \theta^{+\alpha}\bar\theta^{+\dot\alpha}\big)
-(\lambda^M\partial_M h^{++\alpha\dot\alpha})\,.
\eea

The linearized form of the transformation law \eqref{Actives2} reads:
\begin{equation}\label{transpG}
    \delta^*_\lambda \hat{\mathcal{H}}^{++}_{(2)} = \delta_\lambda \hat{\mathcal{H}}^{++}_{(2)} = [\mathcal{D}^{++}, \hat{\Lambda}_{(2)}]\,.
\end{equation}
So we conclude that the analyticity preserving  ${\cal N}=2$ supersymmetry invariant differential operator $\hat{\mathcal{H}}^{++}_{(2)}$ has the very simple linearized transformation law
realized as the commutator of the flat harmonic derivative with the differential  operator involving all the superfield gauge parameters which are contracted with the corresponding
partial derivatives. In the case of spins ${\bf s} \geq 3$ there are
no realizations of the gauge groups on the coordinates, so just the active form of gauge transformations proves to be relevant.

\subsubsection{Differential operators for gauge transformations in $\mathcal{N}=2$ spin s higher spin theory} \label{sec:diff-opp-s}
Now we will extend ${\bf s}=2$ construction of $\mathcal{N}=2$ supersymmetry invariant analytic differential operator $\hat{\mathcal{H}}^{++}_{(2)}$ \eqref{Hspin2}
to {\it general higher $\mathcal{N}=2$ spins ${\bf s}$}. It will be the necessary ingredient for construction of interactions of hypermultiplet with higher spins.

First of all, we will define the analytic $\mathcal{N}=2$ invariant differential operator that is a natural extension of the differential operator \eqref{Hspin2} for ${\bf s}=2$:
\begin{equation}\label{G-operator}
    \hat{\mathcal{H}}^{++\alpha(s-2)\dot{\alpha}(s-2)} := h^{++\alpha(s-2)\dot{\alpha}(s-2)M} \partial_{M}\,.
\end{equation}
Here ${\bf s}\geq 2$ and it is assumed that the corresponding spinor indices in $h^{++\alpha(s-2)\dot{\alpha}(s-2)M}$ are always symmetrized with those in $M$, e.g.,
 $h^{++\alpha(s-2)\dot{\alpha}(s-2)\gamma \dot\gamma} = h^{++(\alpha(s-2)\gamma)(\dot{\alpha}(s-2)\dot\gamma)}$.  The
 corresponding operator involving the analytic gauge parameters is constructed as a natural generalization  of \eqref{diffL}:
\begin{equation}\label{Lambda}
    \hat{\Lambda}^{\alpha(s-2)\dot{\alpha}(s-2)} := \lambda^{\alpha(s-2)\dot{\alpha}(s-2)M} \partial_{M}\,,
\end{equation}
with the same convention concerning the spinor indices hidden in the index $M$, e.g., $\lambda^{\alpha(s-2)\dot{\alpha}(s-2)\gamma \dot\gamma} =
\lambda^{(\alpha(s-2)\gamma)(\dot{\alpha}(s-2)\dot\gamma)}$. In these notations, the  gauge transformations \eqref{Gauge_s} are realized
as
\begin{equation}\label{transf}
\delta_\lambda \hat{\mathcal{H}}^{++}_{(s)} =
[\mathcal{D}^{++}, \hat{\Lambda}_{(s)}]\,,
\end{equation}
where
\begin{eqnarray}\label{operHS}
   && \hat{\mathcal{H}}^{++}_{(s)}: =  \hat{\mathcal{H}}^{++\alpha(s-2)\dot{\alpha}(s-2)} \partial^{s-2}_{\alpha(s-2)\dot{\alpha}(s-2)}\,, \nn
  &&  \hat{\Lambda}_{(s)}
    :=\hat{\Lambda}^{\alpha(s-2)\dot{\alpha}(s-2)}\partial^{s-2}_{\alpha(s-2)\dot{\alpha}(s-2)}
\end{eqnarray}
and
\be
\partial^{k}_{\alpha(k)\dot{\alpha}(k)} := \partial_{(\alpha_1\dot{\alpha}_1}\dots\partial_{\alpha_k)\dot{\alpha}_k}\,.
\ee
Note, that here, in contrast to the spin ${\bf 2}$ case \eqref{transpG}, we added $s-2$ vector derivatives $\partial^{s-2}_{\alpha(s-2)\dot{\alpha}(s-2)}$ in \eqref{transf}.
These extra derivatives are necessary both for reproducing the correct transformation properties of the analytic prepotentials from \eqref{transf} and for
ensuring the compatibility with the rigid ${\cal N}=2$ supersymmetry. Indeed, it is easy to check that only the operator  $\hat{\mathcal{H}}^{++}_{(s)}$
is ${\cal N}=2$ supersymmetric, $\delta_\epsilon \hat{\mathcal{H}}^{++}_{(s)} =0$. For ${\bf s} \geq 3$, neither the operator \eqref{G-operator} nor
any its contraction with $\partial_{\alpha\dot\alpha}$, such that some amount of
the free spinor indices is left, are invariant under ${\cal N}=2$ supersymmetry \eqref{TranPart}, \eqref{susy}.

One can also construct a useful ${\cal N}=2$ {\it superfield}:
\begin{equation}\label{Gamma-main}
    \Gamma^{++}_{(s)} =  \partial^{s-2}_{\alpha(s-2)\dot\alpha(s-2)}    \Gamma^{++\alpha(s-2)\dot{\alpha}(s-2)}\,,
\end{equation}
where we used the  notation
\begin{eqnarray}
   && \Gamma^{++\alpha(s-2)\dot{\alpha}(s-2)} = \partial_{\beta\dot{\beta}}h^{++(\alpha(s-2)\beta)(\dot{\alpha}(s-2)\dot{\beta})}-
    \partial^-_{\beta}h^{++(\alpha(s-2)\beta)\dot{\alpha}(s-2)+} \nonumber \\
    && -  \partial^-_{\dot{\beta}} h^{++\alpha(s-2)(\dot{\alpha}(s-2)\dot{\beta})+} = (-1)^{P(M)} \partial_M h^{++\alpha(s-2)\dot{\alpha}(s-2) M}\,,
\end{eqnarray}
with  $P(M) = 0$ for $M=\alpha\dot\alpha$ and $P(M) = 1$ for $M = \hat\mu +$.
It possesses the simple gauge transformation law:
\begin{equation}\label{Gamma}
    \delta_\lambda \Gamma^{++}_{(s)} =  \mathcal{D}^{++}\Omega_{(s)}\,, \quad \Omega_{(s)} := \big(\partial^{s-2}_{\alpha(s-2)\dot{\alpha}(s-2)}\Omega^{\alpha(s-2)\dot{\alpha}(s-2)}\big),
\end{equation}
where
\be \label{Omega}
\Omega^{\alpha(s-2)\dot{\alpha}(s-2)} := (-1)^{P(M)} \big(\partial_M \lambda^{\alpha(s-2)\dot{\alpha}(s-2) M}\big)\,.
 \ee
The superfield  $\Gamma^{++}_{(s)}$ is also invariant under $\mathcal{N}=2$ supersymmetry \eqref{TranPart} and \eqref{susy}:
\begin{equation}
    \delta_{\epsilon} \Gamma^{++}_{(s)} = 0\,.
\end{equation}

We will use the operator \eqref{operHS} and the superfield \eqref{Gamma-main} for construction of gauge-invariant couplings with the hypermultiplet.
Note that there are only one $\mathcal{N}=2$ supersymmetric invariant operator $\hat{\mathcal{H}}^{++}_{(s)}$ and only one invariant superfield $\Gamma^{++}_{(s)}$
which are linear in the analytic prepotentials. This will significantly restrict the  possible cubic couplings with the hypermultiplet.

\section{Hypermultiplet couplings: spins 1, 2, 3, 4}
\label{sec:hyper}

\subsection{Hypermultiplet in harmonic superspace}
The ${\cal N}=2$ hypermultiplet free action has the
form \cite{18, HSS}\footnote{We use the standard definition of the analytic superspace
        integration measure \cite{18}:
        $$
        d\zeta^{(-4)} := d^4x d^2 \theta^+ d^2 \bar{\theta}^+ du\,.
        $$}:
\begin{equation}\label{hyper}
    S = \int  d\zeta^{(-4)}  \; \mathcal{L}^{+4}_{free} = -\int d\zeta^{(-4)}  \; \frac{1}{2} q^{+a} \mathcal{D}^{++} q^+_a
    = -\int d\zeta^{(-4)}  \; \tilde{q}^+ \mathcal{D}^{++} q^+, \;\;[q^{+a}] = 1.
\end{equation}
We can write the action in the two equivalent forms: in terms of the pseudo-real analytic superfield $q^{+a}(\zeta)$ with the Pauli-G\"ursey ${\rm SU}(2)_{PG}$ doublet indices ($a=1,2$) or
in terms of the complex superfields  $q^{+}$ and $\tilde{q}^{+}$. These two representations are related by:
\begin{equation}
    q^+_a = (q^+, - \tilde{q}^+),
    \;\;\;\;\;\;\;\;\;\;
    \widetilde{q^+_a}\equiv q^{+a} = \epsilon^{ab} q^+_b = (\tilde{q}^+, q^+)\,.
\end{equation}
Hereafter, we will use the first form of the action since the manifest ${\rm SU(2)}_{PG}$ symmetry
crucially simplifies the calculations.

A peculiarity of our theory is the presence of the derivative $\partial_5$ and of the corresponding gauge superfields.
The option $\partial_5 q^{+a} \neq 0$ is most general (though the choice $\partial_5 q^{+a}=0$ is also admissible).
In order to avoid any integration
over $x^5$, we impose the standard Scherk-Schwarz condition that $\partial_5 q^{+a}$ coincides (up to a phase factor) with the action
of some ${\rm U}(1)_{PG} \subset {\rm SU}(2)_{PG}$. Without loss of generality we assume
\begin{equation}\label{global-1}
        q^+(x, \theta^+, u, x^5) = e^{-im x^5}  q^+(x, \theta^+, u) \quad \Leftrightarrow \quad \partial_5 q^{+a} := im (\tau^3)^a_{\;b} q^{+b},
\end{equation}
\begin{equation}\label{tau3}
    (\tau_3)^a_{\;b}
    =
    \begin{pmatrix}
        1 & 0 \\
        0 & -1
    \end{pmatrix}
    =
    - (\tau_3)^{\;a}_{b},
    \;\;\;\;\;\;\;\;\;
    (\tau_3)_{ab} = \epsilon_{ac}   (\tau_3)^c_{\;b}
    =
    \begin{pmatrix}
        0 & -1 \\
        -1 & 0
    \end{pmatrix}.
\end{equation}
The parameter $m$ is a mass of the hypermultiplet, so we deal with
the massive hypermultiplet in the general case.
It is easy to
check that \be
\partial_5 (q^{+a}\mathcal{D}^{++} q^+_a) = 0\,. \label{x5independence}
\ee

The action \eqref{hyper} is invariant under the rigid $\mathcal{N}=2$ supersymmetry,
\begin{equation}\label{sup-hyp}
    \delta^*_{\epsilon} q^{+a} = -\delta_\epsilon Z^M\partial_M q^{+a}\,,
\end{equation}
where we need to take into account also the transformation of $x^5$ \eqref{Tranfifth} for $m\neq 0$. Because of \eqref{x5independence}, it does not affect the $\mathcal{N}=2$ supersymmetry
transformation of the Lagrangian, though modifies  $\mathcal{N}=2$ transformation of $q^{+a}$. The internal symmetry of the single hypermultiplet action is
${\rm SU}(2)_{PG} \times{\rm SU}(2)_{aut}$ in the massless case ($\partial_5 q^{+a} =0$) and ${\rm U}(1)_{PG} \times{\rm SU}(2)_{aut}$ in the massive case ($\partial_5 q^{+a} \neq 0$).

The equation of motion for the free massive hypermultiplet is :
\begin{equation}
    \mathcal{D}^{++} q^{+a} = 0. \lb{hyperEq}
\end{equation}
On shell, the analytic harmonic superfield $q^{+a}$ is reduced to:
\begin{equation} \label{Massive}
    q^+(\zeta) = f^i u^+_i
    + \theta^{+\alpha} \psi_\alpha
    + \bar{\theta}^+_{\dot{\alpha}} \bar{\kappa}^{\dot{\alpha}}
    + m(\theta^+)^2 f^i u^-_i
    - m(\bar{\theta}^+)^2 f^i u^-_i
    + 2i \theta^+ \sigma^n \bar{\theta}^+ \partial_n f^i u_i^-\,.
\end{equation}
Eq. \eqref{hyperEq} also implies the massive equations of motion for the physical fields. We will not
need their explicit form. The massless hypermultiplet corresponds to setting $m=0$ in \eqref{global-1} and \eqref{Massive}.

In the sequel we will derive the coupling of the hypermultiplet to the gauge supermultiplets of integer spins described in the HSS approach
by analytic superfields \eqref{eq:q:1} and \eqref{eq:q:2}.

\subsection{Guiding principles}
\label{3.2}

Before turning to  our basic subject, we will formulate a few generic {\it \'a priori} restrictions on the structure of possible interactions:
\begin{itemize}
    \item {\it \underline{Analyticity}}. We require that coupling must be analytic because hypermultiplet is described by an analytic superfield and the well known couplings
    of the hypermultiplet to the gauge and supergravity $\mathcal{N}=2$ theories preserve the analyticity. Moreover, the harmonic analyticity is one of the major
    and crucial features of the harmonic superspace approach and it proved its power while having constructed $\mathcal{N}=2$ Fronsdal theory in
    the harmonic superspace \cite{Buchbinder:2021ite}.

    \item {\it \underline{$\mathcal{N}=2$ supersymmetry}}. The manifest rigid ${\cal N}=2$ supersymmetry is a necessary ingredient of our construction and
    it is one of the main general reasons for employing harmonic superspace. Since the hypermultiplet has the ${\cal N}=2$ transformation law \eqref{sup-hyp}
    the gauge higher-spin superfields must also appear in a way preserving  ${\cal N}=2$ supersymmetry. This requirement significantly limits the possible form of
    cubic interaction. Using the results of section \ref{sec:diff-opp-s} the gauge superfields can appear only within two possible terms, the first one involving
    the supersymmetry-invariant differential operator \eqref{operHS}, and the  second one involving the supersymmetry-invariant superfield \eqref{Gamma-main}:
    \begin{subequations}\label{constraints}
        \begin{equation}
            S^{(s)}_{int(1)} = \int d\zeta^{(-4)}\;  A_{ab} q^{+a} \hat{\mathcal{H}}^{++}_{(s)} q^{+b}\,,
        \end{equation}
        \begin{equation}
            S^{(s)}_{int(2)} = \int d\zeta^{(-4)}\; \Gamma^{++}_{(s)}  B_{ab} q^{+a}  q^{+b}\,.
        \end{equation}
    \end{subequations}
    Here $A_{ab}$ and $B_{ab} = B_{ba}$ are some matrices, which in general can break the Pauli-G\"ursey ${\rm SU}(2)_{PG}$ symmetry.
    The possible structure of these matrices is fully specified by the gauge symmetry (which, in turn, is determined by the appropriate global symmetry of the
    $q^{+}$ action through the gauging procedure).

    \item {\it \underline{Gauge invariance}}. The most crucial property is gauge invariance in the leading order in the gauge superfield.

    As usual, the gauge transformations should be defined through gauging of the  global symmetry transformations and so should yield global symmetry
    upon restricting gauge parameters to constant values.

    We know the linearized gauge transformation laws of the basic quantities $\hat{\mathcal{H}}^{++}_{(s)}$ \eqref{transf} and $\Gamma^{++}_{(s)} $ \eqref{Gamma}.
    So in the leading order in the gauge parameters, the gauge  variation of the full action should be composed of the two possible terms:
    \begin{subequations}
        \begin{equation}\label{g1}
            \delta_\lambda S^{(s)}_{int(1)} = \int d\zeta^{(-4)}\;  A_{ab} q^{+a} [\mathcal{D}^{++}, \hat{\Lambda}^{\alpha(s-2)\dot{\alpha}(s-2)} ]
            \partial^{s-2}_{\alpha\dot{\alpha}} q^{+b} \;,
        \end{equation}
        \begin{equation}\label{g2}
            \delta_\xi S^{(s)}_{int(2)} = \int d\zeta^{(-4)} \; \left(  \mathcal{D}^{++} \partial^{s-2}_{\alpha\dot{\alpha}}
            \Omega^{\alpha(s-2)\dot{\alpha}(s-2)} \right) B_{ab}\, q^{+a}  q^{+b} \,.
        \end{equation}
 \end{subequations}
    These variations must be canceled by the appropriate variations of the free hypermultiplet action induced by the higher-spin gauge transformations of the hypermultiplet.

\end{itemize}

Thus, the general strategy for building cubic interactions of the hypermultiplet with the given spin ${\bf s}$ supermultiplet should be the following.
As the starting point, one finds out the appropriate global symmetry of the  free hypermultiplet action realized on the  superfield $q^{+a}$.
Secondly, one considers the most general gauging of this global symmetry and singles out the necessary combinations of the gauge transformations of the hypermultiplet
ensuring cancelation of the gauge superfield variations \eqref{g1} and \eqref{g2}. Finally, from the last two steps, one determines the matrices $A_{ab}$ and $B_{ab}$,
and, as a result, derives the sought cubic couplings.

It is worth emphasizing that the general gauging procedure described above makes it possible to gauge all global symmetries realized on the hypermultiplet.
Thus it is capable to yield all admissible local first-order Noether couplings of the hypermultiplet to higher-spin gauge fields (see the relevant discussion also
in section \ref{sec:summary}).

\subsection{Spin ${\bf 1}$ coupling}

The construction of the vector multiplet coupling to hypermultiplet is well known \cite{18, HSS}. Here
we adapt it to the generic form applicable to higher spins.\\

\noindent{\it \underline{Rigid symmetry}}. The free hypermultiplet action \eqref{hyper} is invariant under
the ${\rm U}(1)$ rigid transformation which can be realized as a ``shift'' with respect to $x^5$ with the generator \eqref{global-1}.
Now we will gauge this global symmetry.\\

\noindent{\it \underline{Gauging}}. The spin 1 gauge transformations read:
\begin{equation}\label{hyp1}
    \delta_\lambda q^{+a} = -\lambda^5 \partial_5 q^{+a},
    \;\;\;\;\;
    \partial_5 q^{+a} := i m (\tau^3)^a_{\;b} q^{+b}\,,
\end{equation}
where $\lambda^5 (\zeta)$ is an arbitrary analytic gauge parameter. The formalism of sect. \eqref{sec:diff-opp-s}  is adapted to this degenerate case as
\bea
&& {\cal D}^{++} \;\Rightarrow \;\mathfrak{D}^{++}_{(1)} =  {\cal D}^{++} + \hat{\cal H}^{++}_{(1)}\,, \;  \hat{\cal H}^{++}_{(1)} = h^{++ 5} \partial_5\,, \lb{changex5} \\
&& \delta_\lambda \hat{\cal H}^{++}_{(1)} = [{\cal D}^{++}, \hat{\Lambda}_{(1)}], \; \hat{\Lambda}_{(1)} = \lambda^5 \partial_5 \; \Rightarrow \;
\delta_\lambda h^{++ 5} = {\cal D}^{++}\lambda^5\,. \lb{gauge1}
\eea

The transformation of the  hypermultiplet action \eqref{hyper} under \eqref{hyp1} reads:
\begin{equation}\label{max1}
    \delta_\lambda \mathcal{L}^{+4}_{free} = i\frac{1}{2}  (D^{++}\lambda^5) q^{+a}\partial_5 q^{+}_a = i m\,\frac{1}{2}  (\tau^3)_{ab} (D^{++}\lambda^5) q^{+a}q^{+b}\,.
    \end{equation}
If $D^{++}\lambda^5 =0$, the gauge transformation \eqref{hyp1} becomes a rigid internal symmetry of the action, in the full agreement with gauging procedure.
In order to make the action \eqref{hyper} invariant under gauge transformations \eqref{hyp1}, one needs just the compensating gauge superfield $ h^{++ 5}(\zeta)$
which is introduced through the substitution \eqref{changex5}
\begin{equation}\label{action1}
    \mathcal{L}^{+4}_{free}\;\;\to\;\; \mathcal{L}^{+4(s=1)}_{gauge}  = \mathcal{L}^{+4}_{free} - \frac{1}{2}q^{+a} \hat{\cal H}^{++}_{(1)} q^+_a
    = \mathcal{L}^{+4}_{free} -i m\,\frac{1}{2}  h^{++5} (\tau^3)_{ab} q^{+a} q^{+b}\,.
\end{equation}
Using the gauge transformations $\delta h^{++5} = \mathcal{D}^{++} \lambda^5$, one can choose the Wess-Zumino gauge for $h^{++5}$ as:
\begin{multline}
    h^{++5}_{WZ} = -2i \theta^+\sigma^m \bar{\theta}^+ A_m - i (\theta^+)^2 \bar{\phi} + i (\bar{\theta}^+)^2 \phi \\+ 4 (\bar{\theta}^+)^2 \theta^{+\alpha} \psi^i_\alpha u^-_i - 4 (\theta^+)^2 \bar{\theta}^+_{\dot{\alpha}} \bar{\psi}^{\dot{\alpha}i} u^-_i + (\theta^+)^2 (\bar{\theta}^+)^2 D^{ij} u^-_i u^-_j\;,
\end{multline}
which yields just the off-shell field content of $\mathcal{N}=2$ Maxwell multiplet. The engineering dimension of $h^{++5}$ is $-1$
and the passing to the gauge superfield $h^{++}$ with the zero canonical dimension is accomplished as $h^{++} = m h^{++5}$.
The gauge invariant and ${\cal N}=2$ supersymmetric action of the spin ${\bf 1}$ gauge superfield has the standard form,
\be
S_{(s=1)} = \frac{1}{\kappa_{s=1}^2} \int d^4 x d^8\theta du \, h^{++} h^{--}\,, \qquad [\kappa_{s=1}] = 0\,.  \lb{Spn1Act}
\ee

Note that the action \eqref{action1} is exactly invariant under the gauge transformations \eqref{hyp1} and \eqref{gauge1} in consequence of the relation
\begin{equation}\label{max-inc}
    \delta_\lambda \left((\tau^3)_{ab} q^{+a} q^{+b}\right) \sim 2(\tau^3)_{ab}(\tau_3)^a_c q^{+c} q^{+b} \sim \varepsilon_{bc} q^{+c} q^{+b} = 0.
\end{equation}
In the case of higher spin couplings (${\bf s}>2$), no such an exact gauge invariance is present in the leading order in gauge superfields in view of
absence of the analogous relations. So it is the peculiarity of the spin ${\bf 1}$ coupling only.

 One more notable feature of the simplest case considered is the following.
The relations (3.11) and (3.15) imply that if the hypermultiplet is $x^5$
independent, the mass parameter $m$ vanishes in \eqref{action1} and a non-trivial cubic
coupling of such a massless hypermultiplet to spin ${\bf 1}$ gauge field is seemingly impossible. To avoid
this obstacle and get a possibility to construct the cubic coupling
for both massive and massless theory, one can resort to the following reasoning.
The main point is that in general it is not obligatory to identify the central
charge $\partial_5$ with the generator $J$ of  ${\rm U}(1)_{PG} \subset {\rm
SU}(2)_{PG}$ to be gauged. The basic condition  for such a generator is
\be
[\mathfrak{D}^{++}_{(1)}, J] = 0\,, \lb{CondJ}
\ee
whence, without loss of generality,
\be
J q^{+a} = i  (\tau_3)^a_{\;b} q^{+b}\,.\lb{Part6}
\ee
This corresponds just to replacing $\partial_5$
to $J$ in the gauge transformation law \eqref{hyp1}:
\be
\delta'_\lambda q^{+a} = -\lambda J q^{+a} = -i \,\lambda(\tau_3)^a_{\;b} q^{+b}\,, \quad [\lambda] = 0\,.\label{hyp1mod}
\ee
Hence the hypermultiplet can still be chosen $\partial_5$-neutral ($\partial_5 q^{+a} =0$) and thus massless.
The interaction with Maxwell multiplet will have the same form as in \eqref{action1}, where
we have to replace $m$ by $1$ and $h^{++ 5}$ by the dimensionless  $h^{++}$.
Note that in the {\it massive} case, with $\partial_5 q^{+a} \sim i (\tau_3)^a_{\;b}
q^{+b} \neq 0$, it is {\it necessary} to identify $\partial_5$ with the
gauged ${\rm U}(1)_{PG}$ generator $J$ (up to an unessential
constant of mass dimension, $\partial_5 \sim m J$), in order to preserve the gauge invariance of the total
$q^{+a}$ action. In the {\it massless} case $J$ still commutes with all ${\cal N}=2$ supersymmetry generators (as ${\rm SU}(2)_{PG}$ generators do)
but has  no any relation with the central charge.

Let us point out once more that the ${\cal N}=2$ Maxwell theory analytic potential
and its coupling to the charged hypermultiplet can be constructed without any reference to the fictitious coordinate $x^5$ and its interplay with the generator $J$.
The way we have followed here is just a simple illustration of
the general procedure outlined in the previous subsection. As we will see soon (in the subsections \ref{sec:35} and \ref{sec:41}), the above procedure with
the extra coordinate $x^5$ and the matrix generator $J$ perfectly well works for all odd spins ${\bf s} \geq 3\,$.

We finish this subsection with the two comments.

First, in the gauge-covariant derivative \eqref{changex5} one can remove the mass-generating background in ${\cal D}^{++}$ just
by redefining $h^{++5} \rightarrow h^{++5} - i[(\theta^+)^2 - (\bar\theta^+)^2]$, before any identification of $\partial_5$ with ${\rm U}(1)_{PG} \subset {\rm SU}(2)_{PG}$
\footnote{ Such a redefinition leaves invariant, up to a total harmonic derivative, the action of $h^{++5}$  and slightly changes the ${\cal N}=2$ supersymmetry transformation
of $h^{++5}$ by adding a particular gauge transformation.}.
This means that, in the gauged ${\bf s} =1$ theory for
the single hypermultiplet, there is no actual difference between massive and massless cases: the mass parameter $m$ appears only as a coupling constant in the minimal interaction
\eqref{action1} and can be removed (or set equal to 1) by the proper rescaling of $h^{++ 5}$ (the same phenomenon can be easily traced also in the component formulation).

Secondly, perhaps a more convincing explanation why it is useful to introduce the matrix generator $J \subset su(2)_{PG}$ besides $\partial_5$ is as follows.
As we saw in the previous section, the
differential operators relevant to the spin ${\bf s}$ case are always of the order $({\bf s} -1)$, so it would be more natural to describe the spin \textbf{1}
by the differential operators
of the zeroth order, than of the first order as in \eqref{gauge1}.  Such a description, in accord with the general scheme of Section \ref{sec:hs}, amounts to
\bea
&& {\cal D}^{++} \;\Rightarrow \;\mathfrak{D}^{++}_{(1)} =  {\cal D}^{++} + \hat{\cal H}^{++}\,, \quad  \hat{\cal H}^{++}= h^{++}J\,, \lb{changex56} \\
&& \delta_\lambda \hat{\cal H}^{++} = [{\cal D}^{++}, \hat{\Lambda}]\,, \quad \hat{\Lambda}= \lambda J\; \Rightarrow \;
\delta_\lambda h^{++} = {\cal D}^{++}\lambda\,. \lb{gauge16}
\eea
The flat harmonic derivative ${\cal D}^{++}$ is still defined by the general expression \eqref{Dflat+}, with the central charge operator $\partial_5$.
The relevant gauge transformation of $q^{+a}$ is now {\it postulated} as in \eqref{hyp1mod}, with the ${\rm SU}(2)_{PG}$ symmetry matrix generator $J$ properly realized on $q^{+a}$.
The covariantization of \eqref{changex56}, when it acts on $q^{+a}$, is then accomplished as
\be
\mathfrak{D}^{++}_{(1)} q^{+ a} = \big({\cal D}^{++} + \hat{\cal H}^{++}\big) q^{+ a} = {\cal D}^{++}q^{+ a} + i h^{++}(\tau_3)^a_{\;b} q^{+b}\,.
\ee
The standard condition $[\mathfrak{D}^{++}_{(1)}, J]q^{+ a} = 0$ still admits the solution $\partial_5q^{+ a} = 0\,,$ which yields a massless hypermultiplet.
For $\partial_5q^{+ a} \neq 0\,,$ the same condition necessitates the relation $\partial_5 \sim J$, yielding the massive hypermultiplet. As we argued above, for the spin ${\bf 1}$
coupled to the hypermultiplet,
the massive and the massless  Lagrangians are related through a redefinition of the gauge superfield $h^{++}$. However, no such an equivalence
between massive and massless Lagrangians is valid for higher odd spins ${\bf s} \geq 3$.
It is just the description outlined here that directly extends to the ${\bf s} \geq 3$ case.

\subsection{Spin ${\bf 2}$ coupling}
The hypermultiplet coupling to ${\cal N}=2$ supergravity ({\bf s}=2)  is also well known \cite{HSS, Galperin:1987em, Galperin:1987ek}.
Here we repeat the relevant construction in the form most convenient for further generalizations.\\

\noindent{\it \underline{Rigid symmetry}}. The free hypermultiplet
possesses the following global supetranslational symmetry
\begin{equation}\label{global-2}
    \delta_{rig} q^{+a} =  -\hat{\Lambda}_{rig} q^{+a},
\end{equation}
\begin{eqnarray}
    \hat{\Lambda}_{rig} &=& \left( \lambda^{\alpha\dot{\alpha}} - 2i \lambda^{-\alpha} \bar{\theta}^{+\dot{\alpha}} - 2i \theta^{+\alpha} \bar{\lambda}^{-\dot{\alpha}}  \right) \partial_{\alpha\dot{\alpha}}
    + \lambda^{+\alpha} \partial^-_{\alpha} + \bar{\lambda}^{+\dot{\alpha}} \partial^-_{\dot{\alpha}} \nonumber \\
   &&+\, \left( \lambda^5 + 2i \lambda^{\hat{\alpha}-} \theta^+_{\hat{\alpha}} \right) \partial_5  := \Lambda^M \partial_M\,.
\end{eqnarray}
It involves five constant bosonic parameters $\lambda^{\alpha\dot{\alpha}}$, $\lambda^5$, four constant spinor parameters $\lambda^{\hat{\alpha}i}$,
such that $\lambda^{\pm\hat{\alpha}} = \lambda^{\hat{\alpha}i} u^\pm_i$,
and it can be treated as a copy of the rigid ${\cal N}=2$ supersymmetry transformations in their active form. However, we will gauge just it, leaving ${\cal N}=2$ supersymmetry still rigid,
so that the latter forms a semi-direct product with the gauge extension  of \eqref{global-2}.
Recall that in the previous subsection we have already introduced $\lambda^5$ transformations in order to describe spin ${\bf 1}$ supermultiplet after their gauging.
The symmetry \eqref{global-2} is an extension of this $\partial_5$ symmetry, such that its gauging generates the multiplet
of minimal  $\mathcal{N}=2$ Einstein supergravity.

Operator $\hat{\Lambda}_{rig}$ commutes with the harmonic derivative (we assume, as earlier, that all operators act on analytic superfields):
\begin{equation}
    [\mathcal{D}^{++}, \hat{\Lambda}_{rig}] = 0\,.
\end{equation}
The Lagrangian \eqref{hyper} is invariant, up to total derivative, under these transformations:
\begin{equation}
    \begin{split}
        \delta_{rig} \mathcal{L}^{+4}_{free}
        &=  \frac{1}{2} \hat{\Lambda}_{rig} q^{+a} \mathcal{D}^{++} q^+_a
        +
         \frac{1}{2}  q^{+a} \mathcal{D}^{++} \hat{\Lambda}_{rig} q^+_a
        =
         \frac{1}{2} \hat{\Lambda}_{rig} \left( q^{+a} \mathcal{D}^{++} q^+_a \right)
        \\&=
         (-1)^{P(M)} \frac{1}{2} \partial_M \Lambda^M \left( q^{+a} \mathcal{D}^{++} q^+_a \right) = 0\,.
    \end{split}
\end{equation}

\noindent{\it \underline{Gauging}}. In this case there are two possibilities for gauge transformations of the hypermultiplet:
\begin{equation}\label{q1}
    \delta_1 q^{+a} =  -\hat{\Lambda}_{(2)} q^{+a},
    \;\;\;\;\;
    \hat{\Lambda}_{(2)}: = \lambda^M \partial_M = \lambda^{\alpha\dot{\alpha}} \partial_{\alpha\dot{\alpha}} + \lambda^{+\alpha} \partial^-_{\alpha} + \bar{\lambda}^{+\dot{\alpha}} \partial^-_{\dot{\alpha}} + \lambda^5 \partial_5\,,
\end{equation}
\begin{equation}\label{q2}
    \delta_2 q^{+a} =  -\frac{1}{2} \Omega_{(2)} q^{+a},
    \;\;\;\;\;
    \Omega_{(2)} := (-1)^{P(M)} \partial_M \lambda^M = \partial_{\alpha\dot{\alpha}}\lambda^{\alpha\dot{\alpha}}
     - \partial^-_{\alpha} \lambda^{+\alpha}  - \partial^-_{\dot{\alpha}}\bar{\lambda}^{+\dot{\alpha}}.
\end{equation}
The relevant first- and zeroth-order differential operators  are the particular ${\bf s}=2$ case of the general operators \eqref{Lambda} and \eqref{Omega}.
Here $\lambda^{\alpha\dot{\alpha}}(\zeta), \lambda^{+\hat{\alpha}}(\zeta)$ and $\lambda^5(\zeta)$ are arbitrary analytic gauge parameters.
The first type of transformations, $\delta_1 q^{+a}$, corresponds to the direct gauging of the above supertranslations, while
the second type $\delta_2 q^{+a}$ can be treated as a special gauging of the constant-parameter rescaling of the hypermultiplet (which is not invariance on its own).
Under \eqref{q1} the hypermultiplet action \eqref{hyper}, up to a total derivative, transforms as:
\begin{equation}\label{spin2-1}
    \delta_1 \mathcal{L}^{+4}_{free} =  \frac{1}{2} q^{+a} [\mathcal{D}^{++}, \hat{\Lambda}_{(2)}] q^+_a
    -
     \frac{1}{2} \Omega_{(2)} q^{+a} \mathcal{D}^{++} q^+_a\,.
\end{equation}

The transformation \eqref{q2} leads to the following variation of the Lagrangian:
\begin{equation}
    \delta_2 \mathcal{L}^{+4}_{free} = \frac{1}{2} \Omega_{(2)} q^{+a} \mathcal{D}^{++} q^+_a\,,
\end{equation}
where we made use of the evident property
\begin{equation}
    q^{+a} (\mathcal{D}^{++} \Omega_{(2)}) q^+_a = 0\,.
\end{equation}

Thus the total gauge variation of the free Lagrangian reads:
\begin{equation}
    \left(\delta_1 + \delta_2 \right) \mathcal{L}^{+4}_{free}
    =
     \frac{1}{2} q^{+a} [\mathcal{D}^{++}, \hat{\Lambda}_{(2)}] q^+_a. \lb{TotalVarspin2}
\end{equation}
This variation vanishes for the constant parameters, so the relevant transformations  provide the evident symmetry of the action.

To couple spin ${\bf 2}$ theory to the hypermultiplet we use the differential operator \eqref{Hspin2},
\begin{equation}
    \hat{\mathcal{H}}^{++}_{(2)} = h^{++\alpha\dot{\alpha}} \partial_{\alpha\dot{\alpha}} + h^{++\hat{\mu}+} \partial^-_{\hat{\mu}} + h^{++5}\partial_5\,,
\end{equation}
with the transformation law \eqref{transpG}:
\begin{equation}
    \delta \hat{\mathcal{H}}^{++}_{(2)} = [\mathcal{D}^{++}, \hat{\Lambda}_{(2)}].
\end{equation}
Respectively, the gauging is accomplished as:
\begin{equation}\label{action2}
    \mathcal{L}^{+4}_{free}\;\;\to\;\; \mathcal{L}^{+4 (s=2)}_{gauge} = \mathcal{L}^{+4}_{free} - \frac{1}{2} q^{+a} \hat{\mathcal{H}}^{++}_{(2)} q^+_a\,.
\end{equation}
The variation of $ \hat{\mathcal{H}}^{++}_{(2)}$ in the  second term cancels \eqref{TotalVarspin2}. So the action \eqref{action2} provides
the gauge invariant coupling of the spin ${\bf 2}$ to the hypermultiplet in the leading (first)  order in gauge superfields.

Now we could try to restore the next orders. The $\delta_1$ and $\delta_2$ variations of the hypermultiplet $q^{+a}$ in the second term of \eqref{action2}, up to a total derivative,
generate the following new term:
\begin{equation}
   \tilde\delta \left(\frac{1}{2} q^{+a} \hat{\mathcal{H}}^{++}_{(2)} q^+_a \right) =
    \frac{1}{2} q^{+a} [\hat{\mathcal{H}}^{++}_{(2)}, \hat{\Lambda}_{(2)} ] q^+_a\,.
    \end{equation}
So, for the full  variation of the gauge superfield part  we obtain:
\begin{equation}
   \delta \left(\frac{1}{2} q^{+a} \hat{\mathcal{H}}^{++}_{(2)} q^+_a \right) = -\frac{1}{2} q^{+a} [\mathcal{D}^{++}, \hat{\Lambda}_{(2)}] q^+_a
    +
   \frac{1}{2} q^{+a} [\hat{\mathcal{H}}^{++}_{(2)}, \hat{\Lambda}_{(2)} ] q^+_a\,.
\end{equation}
Thus the action \eqref{action2} is not invariant under the gauge transformations \eqref{q1} and \eqref{q2}. The full gauge transformation
of the Lagrangian \eqref{action2} reads
\be
\delta_\lambda \mathcal{L}^{+4(s=2)}_{gauge} =  \frac{1}{2} q^{+a} [\hat{\mathcal{H}}^{++}_{(2)}, \hat{\Lambda}_{(2)} ] q^+_a\,.
\ee
One can cancel this extra unwanted  term just by adding the nonlinear term to the linearized gauge transformation  law \eqref{transpG}:
\begin{equation}\label{sugra-full}
    \delta_{full} \hat{\mathcal{H}}^{++}_{(2)}  =
    [\mathcal{D}^{++}, \hat{\Lambda}_{(2)}]
    +
    [\hat{\mathcal{H}}^{++}_{(2)}, \hat{\Lambda}_{(2)}] = [\mathfrak{D}^{++}, \hat{\Lambda}_{(2)}] \,.
\end{equation}
Thus, in contrast to the spin ${\bf 1}$ case, where we had the exact gauge invariance without any modification of the transformation law of the relevant operator $\hat{\mathcal{H}}^{++}_{(1)}$,
in the spin ${\bf 2}$ case one is led to modify the gauge transformation of $\hat{\mathcal{H}}^{++}_{(2)}$ to achieve the complete gauge invariance.

The deformation \eqref{sugra-full} of the spin $\mathbf{2}$
transformation law means that the linearized action of
$\mathcal{N}=2$ supergravity \eqref{higher spin action} must also be
modified by including nonlinear terms in the action, so as to achieve the full gauge invariance
for both the pure supergravity gauge superfield action and the hypermultiplet couplings. It is an essential difference from the spin ${\bf 1}$ case
where the pure gauge part of the spin ${\bf 1}$ total action is given by the action \eqref{Spn1Act} and does not require any correction terms. A complete
nonlinear harmonic superfield action for $\mathcal{N}=2$ Einstein supergravity has been
constructed in \cite{Galperin:1987em}. We emphasize that the discussion of any nonlinear
aspects of this kind are beyond the scope of the present paper.

To summarize, we have restored, from the transformation law for hypermultiplet and gauge fields, the well known complete coupling of the hypermultiplet to
the minimal ${\cal N}=2$ supergravity multiplet:

\begin{equation}\label{action3}
    \mathcal{L}^{+4(s=2)}_{gauge} = -\frac{1}{2} q^{+a} \mathfrak{D}^{++} q^+_a
    =
    -\frac{1}{2} q^{+a} \left( \mathcal{D}^{++} + \hat{\mathcal{H}}^{++}_{(2)} \right) q^+_a\,.
\end{equation}

It is worth noting that it is easier to check the gauge invariance of the analytic superspace action with the Lagrangian \eqref{action3}, using the passive form of the gauge
transformations
\be
\delta_\lambda \mathfrak{D}^{++} = 0\,, \quad \delta_\lambda q^{+a} =  -\frac{1}{2} \Omega_{(2)} q^{+a}\,, \quad \delta_\lambda d\zeta^{(-4)} = d\zeta^{(-4)}\Omega_{(2)}\,.
\ee

Our last remark is that it is consistent to choose $\partial_5 q^{+a} = 0$ in the spin ${\bf s}=2$ case. Under this restriction, the  prepotential $h^{++5}$ drops out from
the hypermultiplet action which so corresponds to the massless hypermultiplet. However,  the $h^{++5}$ term should still be present in the gauge action \eqref{ActionsGen} with
${\bf s}=2$ for ensuring the gauge invariance of the latter.

\subsection{Spin ${\bf 3}$ coupling}\label{sec:35}

In this case we will also start with the appropriate {\it \underline{rigid transformations}}. Our previous consideration suggests that for any ${\bf s}$
the $q^{+a}$ variation of the highest order in derivatives involves the same
differential operator as in \eqref{operHS}, $\partial_M \partial^{s-2}_{\alpha(s-2)\dot\alpha(s-2)}$, {\it i.e.} $\partial_M \partial_{\alpha\dot\alpha}$ for ${\bf s}=3$.

Here {\it \'a priory} we have two options for defining higher-order global transformations of $q^{+a}$, such that the second option
(an analog of global transformations of spin ${\bf1}$ \eqref{global-1} in a sense) leads to the desired result, while
the first option (the most natural generalization of the spin ${\bf s} =2$ transformation  \eqref{global-2}) does not ensure
the invariance of the  Lagrangian.\\

\noindent{\it \underline{Option 1}}.
It is chosen as a direct generalization of the spin ${\bf2}$ rigid transformations \eqref{global-2},
\begin{equation}\label{spin3-fal}
    \delta_{rig} q^{+a} = -\hat{\Lambda}_{rig}^{\alpha\dot{\alpha}} \partial_{\alpha\dot{\alpha}} q^{+a}\,,
\end{equation}
\begin{equation}
    \begin{split}
    \hat{\Lambda}_{rig}^{\alpha\dot{\alpha}} =&
    \left(\lambda^{(\alpha\beta)(\dot{\alpha}\dot{\beta})} - 2i \lambda^{(\alpha\beta)(\dot{\alpha}-} \bar{\theta}^{+\dot{\beta})}
    - 2i
    \theta^{+(\alpha}
     \bar{\lambda}^{\beta)(\dot{\alpha}\dot{\beta})-}  \right)\partial_{\beta\dot{\beta}} \\&+ \lambda^{(\alpha\beta)\dot{\alpha}+} \partial^-_\beta
    +
    \bar{\lambda}^{(\dot{\alpha}\dot{\beta})\alpha+} \partial^-_{\dot{\beta}}
    +
     \left( \lambda^{\alpha\dot{\alpha}}
    + 2i \lambda^{(\alpha\beta)\dot{\alpha}-} \theta^+_{{\beta}} + 2i \lambda^{\alpha(\dot\beta\dot{\alpha})-} \bar\theta^+_{{\dot\beta}} \right) \partial_5\,.
        \end{split}
\end{equation}

Here we used the notation \eqref{Lambda} for the differential operator $\hat{\Lambda}^{\alpha\dot{\alpha}}$ in the ${\bf s}=3$ case. The bosonic parameters
$\lambda^{(\alpha\beta)(\dot{\alpha}\dot{\beta})}, \lambda^{\alpha\dot{\alpha}}$, as well as the coefficients of the harmonics in
the fermionic parameters $\lambda^{(\alpha\beta)\dot{\alpha}\pm} = \lambda^{(\alpha\beta)\dot{\alpha}i}u_i^\pm\,,\;
\bar{\lambda}^{(\dot{\alpha}\dot{\beta})\alpha\pm} = \bar{\lambda}^{(\dot{\alpha}\dot{\beta})\alpha i} u^\pm_i,$
are assumed to be coordinate-independent. So we face $9+4=13$ bosonic parameters and $2\cdot 3 \cdot 2 = 12$ fermionic parameters.
Assuming that the operators act on the analytic superfields, we check that:
\begin{equation}
    [\mathcal{D}^{++}, \hat{\Lambda}_{rig}^{\alpha\dot{\alpha}}] \partial_{\alpha\dot{\alpha}} = 0.\label{Comm3}
\end{equation}
Using this property and integration by parts one can derive the relevant variation of the free Lagrangian \footnote{Hereafter, we always omit total derivatives while operating
with the variations of the Lagrangians.}:
\begin{equation}\label{delta1}
    \begin{split}
        \delta_{rig} \mathcal{L}^{+4}_{free} =\,&  \frac{1}{2} \epsilon_{ab} \left( \hat{\Lambda}_{gl}^{\alpha\dot{\alpha}}
        \partial_{\alpha\dot{\alpha}} q^{+a} \mathcal{D}^{++}q^{+b} +  q^{+a} \mathcal{D}^{++} \hat{\Lambda}_{gl}^{\alpha\dot{\alpha}} \partial_{\alpha\dot{\alpha}} q^{+b}   \right)
        \\=\,&
         \hat{\Lambda}_{rig}^{\alpha\dot{\alpha}} \partial_{\alpha\dot{\alpha}} q^{+a} \mathcal{D}^{++}q^{+}_a\,.
    \end{split}
\end{equation}
We observe that the variation of the Lagrangian is not reduced to a total derivative, which means that the transformations \eqref{spin3-fal}
do not constitute a global symmetry of the free hypermultiplet. Thus we need to look for some alternative definition of the rigid transformations
of the hypermultiplet in the spin ${\bf s}=3$ case.\\

\noindent{\it \underline{Option 2}}. This is a generalization of the spin ${\bf 1}$  rigid transformations \eqref{global-1}.
We take the rigid transformations in the form:
\begin{equation}\label{in3}
    \delta_{rig} q^{+a} = -  \hat{\Lambda}_{rig}^{\alpha\dot{\alpha}} \partial_{\alpha\dot{\alpha}} (J q^{+a}) =
    - i  (\tau^3)^a_{\;b} \hat{\Lambda}_{rig}^{\alpha\dot{\alpha}}  \partial_{\alpha\dot{\alpha}} q^{+b}.
\end{equation}
As in the spin ${\bf s} = 1$ case,  we have introduced here a new generator $J q^{+a}$ which  {\it \'a priori} is not obliged
to be collinear to $\partial_5$ appearing in ${\cal D}^{++}$. To substantiate the choice \eqref{in3}, we note that the relevant modification of the important condition \eqref{Comm3},
\be
[\mathcal{D}^{++}, \hat{\Lambda}_{rig}^{\alpha\dot{\alpha}}J] \partial_{\alpha\dot{\alpha}} = 0\,, \lb{Comm3mod}
\ee
when applied to the hypermultiplet superfield $q^{+a}$, implies, apart from \eqref{Comm3}, also the condition
\be
[\mathcal{D}^{++}, J] = 0\;. \lb{Comm36}
\ee
Assuming that $J$ is identified with some ${\rm U}(1)_{PG} \subset {\rm SU}(2)_{PG}$, like $\partial_5$, eq. \eqref{Comm36} leaves us
with only two possibilities:
\be
({\rm a}) \;\; \partial_5 q^{+ a} = 0\,, \; J q^{+ a} \neq 0\,; \quad ({\rm b}) \;\; \partial_5 q^{+ a}  \sim J q^{+ a} = i  (\tau^3)^{a}_{\;b} q^{+ b} \,,
\lb{SolComm36}
\ee
the second option being related to the fact that no mutually commuting two generators can be found in ${\rm SU}(2)$. Without loss of generality,
$J q^{+ a}$ in (\ref{SolComm36}a) can be chosen the same as in (\ref{SolComm36}b).

The variation of the free Lagrangian under \eqref{in3} is vanishing,
\begin{equation}
    \delta_{rig} \mathcal{L}^{+4}_{free} = -i  \frac{1}{2} (\tau^3)_{ab} \hat{\Lambda}_{rig}^{\alpha\dot{\alpha}} \partial_{\alpha\dot{\alpha}} q^{+b} \mathcal{D}^{++} q^{+a}
    +i
    \frac{1}{2}  (\tau^3)_{ab} q^{+a} \mathcal{D}^{++} \hat{\Lambda}_{rig}^{\alpha\dot{\alpha}} \partial_{\alpha\dot{\alpha}} q^{+b} =0\,,
\end{equation}
where we used \eqref{Comm3mod} and integrated by parts, with taking account of the properties $\partial_M {\Lambda}_{rig}^{\alpha\dot{\alpha}M} =
\partial_{\beta\dot\beta}{\Lambda}_{rig}^{\alpha\dot{\alpha}M} =0\,, \;(\tau^3)_{ab} = (\tau^3)_{ba}\,.$
So the transformation \eqref{in3} provides a global symmetry of the free hypermultiplet action. Thus we are led to gauge just this symmetry.  \\

\noindent{\it \underline{Gauging}}. The gauge transformation operators in this case are
\bea
&& \hat{\Lambda}^{\alpha\dot\alpha} =\lambda^{\alpha\dot\alpha M}\partial_M\,, \quad \hat{\Lambda}_{(3)} = \hat{\Lambda}^{\alpha\dot\alpha}\partial_{\alpha\dot\alpha} =
\lambda^{\alpha\dot\alpha M}\partial_M\partial_{\alpha\dot\alpha}\,, \nonumber \\
&& \Omega^{\alpha\dot\alpha} = (-1)^{P(M)}(\partial_M\lambda^{\alpha\dot\alpha M})\,, \quad \Omega_{(3)} = (\partial_{\alpha\dot\alpha}\Omega_{(3)}^{\alpha\dot\alpha})\,,  \lb{Def3}
\eea
where
\be
\lambda^{\alpha\dot\alpha M}\partial_M = \lambda^{(\alpha\beta)(\dot{\alpha}\dot{\beta})}\partial_{\beta\dot{\beta}} + \lambda^{(\alpha\beta)\dot{\alpha}+} \partial^-_\beta +
    \bar{\lambda}^{(\dot{\alpha}\dot{\beta})\alpha+} \partial^-_{\dot{\beta}} + \lambda^{\alpha\dot{\alpha}} \partial_5\,,
\ee
and all gauge parameters are unconstrained analytic superfields.

There are few admissible transformations laws, which generalize global symmetry \eqref{in3}. It is convenient to choose  the following ones as independent:
    \begin{eqnarray}
 && \delta_{1} q^{+a} = -i\hat{\Lambda}^{\alpha\dot{\alpha}} \partial_{\alpha\dot{\alpha}} (J q^{+a}) = -i\lambda^{\alpha\dot\alpha M}\partial_M\partial_{\alpha\dot\alpha}
       (J q^{+a})\,,\label{gauge-gen-3b} \\
       && \delta_{2} q^{+a} = -i\partial_{\alpha\dot{\alpha}} \hat{\Lambda}^{\alpha\dot{\alpha}}  J q^{+a}
        = -i (\partial_{\alpha\dot{\alpha}}\lambda^{\alpha\dot\alpha M})\partial_M (J q^{+a}) + \delta_{1} q^{+a}\,,\label{gauge-gen-3a} \\
       && \delta_{3} q^{+a} = -i\partial_{\alpha\dot{\alpha}} \Omega^{\alpha\dot{\alpha}} J q^{+a}
        = -i[\Omega_{(3)}
        + \Omega^{\alpha\dot{\alpha}}\partial_{\alpha\dot{\alpha}}]J q^{+a}\,,\label{gauge-gen-3c} \\
      &&  \delta_{4} q^{+a} = -i \Omega^{\alpha\dot{\alpha}} \partial_{\alpha\dot{\alpha}} (J q^{+a})\,.\label{gauge-gen-3d}
    \end{eqnarray}

We vary ${\cal L}^{+4}_{free}$  by the transformations  \eqref{gauge-gen-3b} - \eqref{gauge-gen-3d} with arbitrary coefficients (up to a common rescaling),
using the integration by parts and the relations like
\bea
&& (\tau^3)_{ab}q^{+a}{\cal D}^{++} q^{+b} = \frac12(\tau^3)_{ab}{\cal D}^{++} (q^{+a} q^{+b})\,, \nonumber \\
&&  (\tau^3)_{ab}q^{+a}\partial_{\alpha\dot\alpha}q^{+b}
= \frac12(\tau^3)_{ab} \partial_{\alpha\dot\alpha} (q^{+a} q^{+b})\,. \nonumber
\eea
In particular, up to a total derivative,
\bea
&& \delta_1 {\cal L }^{+4}_{free} = i\frac12 (\tau_3)_{ab} q^{+a}[{\cal D}^{++},\hat{\Lambda}^{\alpha\dot\alpha}] \partial_{\alpha\dot\alpha} q^{+b} \nn
&&+ i \,\frac12 (\tau_3)_{ab}\big[(\partial_{\alpha\dot\alpha} \Lambda^{\alpha\dot\alpha M})\partial_Mq^{+a} {\cal D}^{++}q^{+b} - \Omega^{\alpha\dot\alpha}q^{+a}
\partial_{\alpha\dot\alpha}{\cal D}^{++} q^{+b}\big]. \nonumber
\eea

Then, properly fixing the numerical coefficients, we  single out two appropriate combinations of the variations \eqref{gauge-gen-3b} - \eqref{gauge-gen-3d},
\begin{equation}
    \delta_{\lambda} q^{+a} = \frac12\left( \delta_1 + \delta_2 + \delta_3 \right)q^{+ a}\,,
\end{equation}
\begin{equation}
    \delta_{\xi} q^{+a} := \xi\left( \delta_3  - \delta_4 \right)q^{+a}\,,
\end{equation}
such that they have the necessary form \eqref{g1}, \eqref{g2}
for ${\bf s}=3$:
\begin{equation}\label{transfH3}
    \delta_{\lambda} \mathcal{L}^{+4}_{free} = \frac12\left( \delta_1 + \delta_2 + \delta_3 \right) \mathcal{L}^{+4}_{free}
    =
    i\frac12(\tau^3)_{ab}  q^{+a}  [\mathcal{D}^{++}, \hat{\Lambda}^{\alpha\dot{\alpha}}] \partial_{\alpha\dot{\alpha}} q^{+b}\,,
\end{equation}
\begin{equation}\label{transfH31}
    \delta_{\xi} \mathcal{L}^{+4}_{free} = \xi(\delta_3 - \delta_4) \mathcal{L}^{+4}_{free}
    =
    i\frac{\xi}{2} \left(\mathcal{D}^{++}  \Omega_{(3)}\right) (\tau^3)_{ab}  q^{+a}  q^{+b}\,,
\end{equation}
$\xi$ being some real parameter. Thus, the requirement of gauge invariance has drastically limited the possible form of admissible gauge transformations of the hypermultiplet.

Using the transformation law of the gauge operator $\mathcal{H}^{++}_{(3)}$ \eqref{transf},
\begin{equation}
    \delta \hat{\mathcal{H}}^{++}_{(3)} = \delta \hat{\mathcal{H}}^{++\alpha\dot{\alpha}} \partial_{\alpha\dot{\alpha}} =
    [\mathcal{D}^{++}, \hat{\Lambda}^{\alpha\dot{\alpha}}] \partial_{\alpha\dot{\alpha}}\,,
\end{equation}
where
\begin{equation}
    \hat{\mathcal{H}}^{++\alpha\dot{\alpha}} = h^{++(\alpha\beta)(\dot{\alpha}\dot{\beta})}\partial_{\beta\dot{\beta}} + h^{++(\alpha\beta)\dot{\alpha}+} \partial^-_\beta
    +
    \bar{h}^{++(\dot{\alpha}\dot{\beta})\alpha+} \partial^-_{\dot{\beta}} + h^{++\alpha\dot{\alpha}} \partial_5\,,
\end{equation}
as well as the transformation law for the superfield $\Gamma^{++}_{(3)}$,
\begin{equation}
    \delta \Gamma^{++}_{(3)} = \partial_{\alpha\dot{\alpha}}\delta \Gamma^{++\alpha\dot{\alpha}}
    =
    \mathcal{D}^{++} \Omega_{(3)}\,,
\end{equation}
where
\begin{equation}
    \Gamma^{++\alpha\dot{\alpha}} = \partial_{\beta\dot{\beta}}h^{++(\alpha\beta)(\dot{\alpha}\dot{\beta})}
    -
    \partial^-_{\beta}h^{++(\alpha\beta)\dot{\alpha}+}
    -
    \partial^-_{\dot{\beta}}
    h^{++\alpha(\dot{\alpha}\dot{\beta})+}  \,,
\end{equation}
we can cancel the remainder \eqref{transfH3} and \eqref{transfH31} of the gauge variation of the free Lagrangian by introducing {\it couplings to the gauge superfields} as
\begin{equation}\label{spin3-gauge}
    \begin{split}
        \mathcal{L}^{+4(s=3)}_{gauge} =\,& \mathcal{L}^{+4}_{free}
        - i
          \frac12(\tau^3)_{ab}   q^{+a}  \hat{\mathcal{H}}^{++\alpha\dot{\alpha}}   \partial_{\alpha\dot{\alpha}} q^{+b}
        +
        i
        \frac{\xi}{2} \Gamma^{++}_{(3)} (\tau^3)_{ab}  q^{+a}  q^{+b}
        \\=& -\frac12q^{+a} \left(\mathcal{D}^{++} + \hat{\mathcal{H}}^{++}_{(3)} J +\xi \Gamma^{++}_{(3)} J  \right) q^+_a\,.
    \end{split}
\end{equation}

It will be also useful for the future generalizations to rewrite the transformation laws $\delta_{\lambda} q^{+a}$ and $\delta_{\xi} q^{+a}$ as:
\begin{eqnarray}\label{tranf-spin3-final}
    \delta_{\lambda} q^{+a} &&=
    - \frac12\{\partial_{\alpha\dot{\alpha}},  \hat{\Lambda}^{\alpha\dot{\alpha}}  \} J q^{+a}
    -\frac12\Omega_{(3)}  J q^{+a} 
    -\frac12 \Omega^{\alpha\dot{\alpha}}\partial_{\alpha\dot{\alpha}} J q^{+a} 
    \nn
&&=\, -\left[ \lambda^{\alpha\dot\alpha M}\partial_M\partial_{\alpha\dot\alpha} + \frac12 ( \partial_{\alpha\dot\alpha}\lambda^{\alpha\dot\alpha M})\partial_M
+ \frac12\Omega_{(3)}
+ \frac12 \Omega^{\alpha\dot{\alpha}}\partial_{\alpha\dot{\alpha}}\right] (\tau_3)^a_{\;b} q^{+ b}\,,
\end{eqnarray}

\begin{equation}\label{xi}
\delta_{\xi} q^{+a} =  - \xi \,\Omega_{(3)}\,J q^{+a} = - i\xi \,\Omega_{(3)}\,(\tau_3)^a_{\;b} q^{+ b}\,,
\end{equation}
where in \eqref{tranf-spin3-final} all derivatives act freely on the right.

The obtained gauge-invariant action \eqref{spin3-gauge} demonstrates a freedom in constructing interactions in the ${\bf s}=3$ case: it contains an arbitrary parameter $\xi$.
Note that the transformation \eqref{xi} formally coincides with the gauge transformation for ${\bf s}=1$ \eqref{hyp1} with the parameter
$\partial_{\alpha\dot{\alpha}} \Omega^{\alpha\dot{\alpha}}$. So the transformation \eqref{xi} is the spin ${\bf 1}$ gauge transformation
with the gauge parameter of special form. An analogous ``$\xi$-freedom'' takes place for all odd spins.
The presence of constant $\xi$ in the Lagrangian \eqref{spin3-gauge} shows that off shell there are 2 types of possible interactions of the ${\cal N}=2$ spin ${\bf 3}$
with the hypermultiplet. The coefficient $\xi$ is a dimensionless coupling constant that measures the relative strength of these interactions.

The action \eqref{spin3-gauge} is invariant only up to the leading order in the gauge fields.
In contrast to the cases of spin ${\bf s}=1$, where the action \eqref{action1} was completely invariant, and of spin ${\bf s}=2$, where the coupling \eqref{action3} can  be made invariant
by adding extra terms to the transformation law of gauge superfields, in the case of spin ${\bf s}=3$
some other mechanisms (if exist) are needed\footnote{Perhaps, the full invariance could be achieved after extending the standard $4D,\; {\cal N}=2$ superspace by some additional
coordinates \cite{Buchbinder:2021ite}.}.
In the present paper we limit ourselves to the invariances only in the leading order in gauge superfields.\\

\noindent{\it \underline{Last but not least}}. As we saw, the consistent minimal coupling of the hypermultiplet to the gauge ${\cal N}=2$ spin ${\bf s}=3$ superfields
is possible only  provided $J q^{+a} = i (\tau^3)^a_{\;b} q^{+ b} \neq 0$, which implies that this coupling necessarily breaks ${\rm SU}(2)_{PG}$ down to ${\rm U}(1)_{PG}$\,,
the generator of which is further identified with $J$. In accord with \eqref{SolComm36}, the hypermultiplet can still stay massless [option $({\rm a})$, with $\partial_5q^{+a} = 0$],
or massive [option $({\rm b})$], such that the operator $J$ is proportional to the central charge $\partial_5$ which is not vanishing in the second case. These two possibilities are essentially
different because setting $\partial_5 q^{+a} = 0$ in the second case not only makes the hypermultiplet massless but also eliminates all its couplings to $
{\cal N}=2$ spin ${\bf s}=3$ superfields. As distinct from the simplest ${\bf s}=1$ case, in the ${\bf s}=3$ case it is impossible to relate the massless and massive
hypermultiplet Lagrangians  by any redefinition of the gauge ${\cal N}=2$ potentials.
The same features are characteristic of all ${\cal N}=2$ odd spins (see below). On the contrary, the hypermultiplet couplings to the higher spin ${\cal N}=2$ spin multiplets
with even ${\bf s}$ can be defined for both massive ($\partial_5 q^{+a}\neq 0$) and massless ($\partial_5 q^{+a} = 0$) cases  on equal footing, without
insertions of the ${\rm SU}(2)_{PG}$ generators in the transformation laws and cubic superfield coupling. Note that the full symmetry of the odd spin ${\bf s}$ case is
always ${\rm U}(1)_{PG}\times {\rm SU}(2)_{aut} $, as distinct from the maximal symmetry ${\rm SU}(2)_{PG}\times {\rm SU}(2)_{aut} $ of the even spin case
(which can be attained for $\partial_5 q^{+a}= 0$).

\subsection{Spin ${\bf 4}$ coupling}

\noindent{\it \underline{Rigid symmetry}}. Similarly to other cases, the rigid symmetry to be gauged is given by the maximal-degree differential operator:
\begin{eqnarray}\label{global4}
\delta^{(4)}_{rig} q^{+a} &=& -\hat{\Lambda}_{rig}^{(\alpha\beta)(\dot{\alpha}\dot{\beta})} \partial_{\alpha\dot{\alpha}}\partial_{\beta\dot{\beta}} q^{+a}\,, \label{Glob4}
\eea
with
\bea
\hat{\Lambda}_{rig}^{(\alpha\beta)(\dot{\alpha}\dot{\beta})}
    &=& \big[ \lambda^{(\alpha\beta\gamma)(\dot{\alpha}\dot{\beta}\dot{\gamma})}
    -
    2i \lambda^{(\alpha\beta\gamma)(\dot{\alpha}\dot{\beta}-} \bar{\theta}^{+\dot{\gamma})}
    -
    2i
    \theta^{+(\alpha}\lambda^{\beta\gamma)(\dot{\alpha}\dot{\beta}\dot{\gamma})+}
    \big]  \partial_{\gamma\dot{\gamma}} \nonumber \\
&& + \,\lambda^{(\alpha\beta\gamma)(\dot{\alpha}\dot{\beta})+} \partial^-_{\gamma}
    +
    \lambda^{(\alpha\beta)(\dot{\alpha}\dot{\beta}\dot{\gamma})+}  \partial^-_{\dot{\gamma}} \nonumber \\
&& +\,
    \big[\lambda^{(\alpha\beta)(\dot{\alpha}\dot{\beta})} + 2i \lambda^{(\alpha\beta\gamma)(\dot{\alpha}\dot{\beta})-} \theta^+_{{\gamma}}  +
    2i \lambda^{(\alpha\beta)(\dot{\alpha}\dot{\beta}\dot\gamma)-} \bar\theta^+_{{\dot\gamma}}\big] \partial_5\nn
&:=&  \Lambda^{(\alpha\beta)(\dot{\alpha}\dot{\beta})M}_{rig} \partial_M\,.\label{glob4lamb}
\end{eqnarray}
Fermionic parameters have the form $\lambda^{(\alpha\beta\gamma)(\dot{\alpha}\dot{\beta})\pm}
= \lambda^{(\alpha\beta\gamma)(\dot{\alpha}\dot{\beta})i} u^{\pm}_i$,
$\lambda^{(\alpha\beta)(\dot{\alpha}\dot{\beta}\dot{\gamma})\pm}  = \lambda^{(\alpha\beta)(\dot{\alpha}\dot{\beta}\dot{\gamma})i} u_i^\pm $.
The coefficient of the harmonic variables in these expressions, equally  as the rest of parameters in \eqref{glob4lamb}, are constants. So one has
total of $16+ 6 = 22$ bosonic parameters and of $2\cdot 4 \cdot 3 = 24$ fermionic parameters. The dependence on analytic $\theta$'s in \eqref{glob4lamb} is necessary for
vanishing of the commutator
\begin{equation}
    [\mathcal{D}^{++},  \hat{\Lambda}_{rig}^{(\alpha\beta)(\dot{\alpha}\dot{\beta})}] \partial_{\alpha\dot{\alpha}} \partial_{\beta\dot{\beta}} = 0\,, \label{Comm4}
\end{equation}
which in the present and other cases is  just the condition of the rigidity of the relevant transformation of $q^{+a}$ (eq. \eqref{global4} in the present case, or
eq. \eqref{Comm3}  in the spin ${\bf s}=3$ case).

The variation of the action ${\cal L}^{+4}_{free}$ under the  transformation \eqref{global4} is easily checked to vanish (modulo total derivatives),
\begin{eqnarray}\label{global-spin4}
    \delta^{(4)}_{rig} \mathcal{L}^{+4}_{free} &&= \frac{1}{2}
    (\hat{\Lambda}_{rig}^{\alpha\beta\dot{\alpha}\dot{\beta}} \partial_{\alpha\dot{\alpha}}\partial_{\beta\dot{\beta}} q^{+a}) \mathcal{D}^{++} q^+_a
    +
    \frac{1}{2}
    q^{+a} \mathcal{D}^{++}     (\hat{\Lambda}^{\alpha\beta\dot{\alpha}\dot{\beta}}_{rig} \partial_{\alpha\dot{\alpha}}\partial_{\beta\dot{\beta}} q^+_a) \nn
  &&  =\,
    \frac{1}{2}
    q^{+a} [\mathcal{D}^{++},\hat{\Lambda}^{\alpha\beta\dot{\alpha}\dot{\beta}}_{rig} ]\partial_{\alpha\dot{\alpha}}\partial_{\beta\dot{\beta}} q^+_a = 0\,,
\end{eqnarray}
where we made use of the condition \eqref{Comm4} and, at the intermediate steps, of the evident properties $\partial_M \hat{\Lambda}^{\alpha\beta \dot\alpha\dot\beta M}_{rig}
=\partial_{\alpha\dot{\alpha}}\hat{\Lambda}^{\alpha\beta \dot\alpha\dot\beta M}_{rig} = 0\,$.

 So we have picked up the appropriate symmetry of the free hypermultiplet action. Now we will gauge it. \\

\noindent{\it \underline{Gauging}}. As before, we promote the constant parameters to arbitrary analytic superfields. One can construct six independent
transformation laws with the parameters $\lambda^M$ and derivatives thereof. It will be convenient to choose as a basis the following gauge variations:
\begin{eqnarray}
&& \delta_1 q^{+a} = -\partial_{\alpha\dot{\alpha}} \partial_{\beta\dot{\beta}} \hat{\Lambda}^{\alpha\beta\dot{\alpha}\dot{\beta}} q^{+a}\,, \label{tr3} \\
&& \delta_2 q^{+a} = -\partial_{\alpha\dot{\alpha}} \hat{\Lambda}^{\alpha\beta \dot{\alpha}\dot{\beta}} \partial_{\beta\dot{\beta}} q^{+a}\,, \label{tr1} \\
&& \delta_3 q^{+a} =  -\hat{\Lambda}^{\alpha\beta\dot{\alpha}\dot{\beta}} \partial_{\alpha\dot{\alpha}} \partial_{\beta\dot{\beta}} q^{+a}\,, \label{tr5} \\
&& \delta_4 q^{+a} = -\partial_{\alpha\dot{\alpha}} \partial_{\beta\dot{\beta}} \Omega^{\alpha\beta\dot{\alpha}\dot{\beta}} q^{+a}\,, \label{tr4}\\
&& \delta_5 q^{+a} = -\partial_{\alpha\dot{\alpha}} \Omega^{\alpha\beta\dot{\alpha}\dot{\beta}} \partial_{\beta\dot{\beta}} q^{+a}\,, \label{tr2} \\
&& \delta_6 q^{+a} =  -\Omega^{\alpha\beta\dot{\alpha}\dot{\beta}} \partial_{\alpha\dot{\alpha}} \partial_{\beta\dot{\beta}} q^{+a}\,. \label{tr6}
\end{eqnarray}
Here we used the general definition \eqref{Lambda} for $\hat{\Lambda}^{\alpha\beta\dot{\alpha}\dot{\beta}}$ and \eqref{Omega} for $\Omega^{\alpha\beta\dot{\alpha}\dot{\beta}}$.
The derivatives act freely on all objects to the right,
in accord with our previous conventions. The crucial difference from the spin ${\bf s}=3$ case is the absence of the derivative $\partial_5$ (or the generator $J$)
in these transformation laws.
As a result, they are non-trivial for both the $\partial_5 q^{+a} =0$ and the $\partial_5 q^{+a} \neq 0$ cases. In fact, they are a generalization of the spin ${\bf s}=2$ gauge
transformations.

The gauge variations of ${\cal L}^{+4}_{free}$ are performed straightforwardly, integrating by parts at the intermediate steps. We present only the final answers
\begin{eqnarray}
&& \delta_1 \mathcal{L}^{+4}_{free} =  \frac12\big[\big( \partial_{\alpha\dot{\alpha}} \partial_{\beta\dot{\beta}} \hat{\Lambda}^{\alpha\beta\dot{\alpha}\dot{\beta}}
        -
        \hat{\Lambda}^{\alpha\beta\dot{\alpha}\dot{\beta}}
        \partial_{\alpha\dot{\alpha}} \partial_{\beta\dot{\beta}}
        -
        \Omega^{\alpha\beta\dot{\alpha}\dot{\beta}}
        \partial_{\alpha\dot{\alpha}} \partial_{\beta\dot{\beta}} \big)q^{+a}\big] {\cal D}^{++} q^+_a \nn
&& +\,\frac{1}{2}    q^{+a} [\mathcal{D}^{++}, \hat{\Lambda}^{\alpha\beta\dot{\alpha}\dot{\beta}}] \partial_{\alpha\dot{\alpha}} \partial_{\beta\dot{\beta}}  q^+_a
        +
        \frac{1}{2}  \left(\mathcal{D}^{++} \Omega^{\alpha\beta\dot{\alpha}\dot{\beta}}\right)  q^{+a}  \partial_{\alpha\dot{\alpha}} \partial_{\beta\dot{\beta}}  q^+_a\,,  \label{var3} \\
&& \delta_2 \mathcal{L}^{+4}_{free} =
        -\frac{1}{2} \big(\partial_{\alpha\dot{\alpha}} \Omega^{\alpha\beta \dot{\alpha}\dot{\beta}} \partial_{\beta\dot{\beta}} q^{+a}\big) \mathcal{D}^{++} q^+_a
        \nn&&
        +\,
        \frac{1}{2}  q^{+a} \partial_{\alpha\dot{\alpha}} [\mathcal{D}^{++}, \hat{\Lambda}^{\alpha\beta\dot{\alpha}\dot{\beta}}] \partial_{\beta\dot{\beta}} q^+_a\,, \label{var1} \\
&& \delta_3 \mathcal{L}^{+4}_{free} = -\frac{1}{2}  \Big[ \left( \partial_{\alpha\dot{\alpha}} \partial_{\beta\dot{\beta}} \hat{\Lambda}^{\alpha\beta\dot{\alpha}\dot{\beta}}
        -
        \hat{\Lambda}^{\alpha\beta\dot{\alpha}\dot{\beta}}
        \partial_{\alpha\dot{\alpha}} \partial_{\beta\dot{\beta}}
        +
        \partial_{\alpha\dot{\alpha}} \partial_{\beta\dot{\beta}}   \Omega^{\alpha\beta\dot{\alpha}\dot{\beta}} \right) q^{+a}\Big] \mathcal{D}^{++} q^+_a \nn
&& +\, \frac{1}{2}
        q^{+a}  [\mathcal{D}^{++}, \hat{\Lambda}^{\alpha\beta\dot{\alpha}\dot{\beta}}] \partial_{\alpha\dot{\alpha}} \partial_{\beta\dot{\beta}} q^+_a\,, \label{var5} \\
&& \delta_4 \mathcal{L}^{+4}_{free} =
        \frac{1}{2}  \Big[\left( \partial_{\alpha\dot{\alpha}} \partial_{\beta\dot{\beta}}  \Omega^{\alpha\beta\dot{\alpha}\dot{\beta}}
        +
        \Omega^{\alpha\beta\dot{\alpha}\dot{\beta}}
        \partial_{\alpha\dot{\alpha}} \partial_{\beta\dot{\beta}}  \right)q^{+a}\Big]
        \mathcal{D}^{++}q^+_a \nn
&& -\,\frac{1}{2}  \left(\mathcal{D}^{++} \Omega^{\alpha\beta\dot{\alpha}\dot{\beta}}\right) q^{+a}   \partial_{\alpha\dot{\alpha}}\partial_{\beta\dot{\beta}} q^+_a\,, \label{var4} \\
&& \delta_5 \mathcal{L}^{+4}_{free} =  \big(\partial_{\alpha\dot{\alpha}} \Omega^{\alpha\beta\dot{\alpha}\dot{\beta}} \partial_{\beta\dot{\beta}} q^{+a}\big)
\mathcal{D}^{++} q^+_a\,,\label{var2} \\
&& \delta_6 \mathcal{L}^{+4}_{free} =
        \frac{1}{2} \Big[\left( \partial_{\alpha\dot{\alpha}} \partial_{\beta\dot{\beta}}  \Omega^{\alpha\beta\dot{\alpha}\dot{\beta}}
        +
        \Omega^{\alpha\beta\dot{\alpha}\dot{\beta}}
        \partial_{\alpha\dot{\alpha}} \partial_{\beta\dot{\beta}}  \right)q^{+a}\Big]
        \mathcal{D}^{++}q^+_a \nn
&& +\, \frac{1}{2} \left(\mathcal{D}^{++} \Omega^{\alpha\beta\dot{\alpha}\dot{\beta}}\right) q^{+a}  \partial_{\alpha\dot{\alpha}}\partial_{\beta\dot{\beta}} q^+_a\,. \label{var6}
\end{eqnarray}

These variations still admit, through integration by parts,  some other forms sometimes more convenient for calculations, e.g.,
\bea
&& \delta_3 \mathcal{L}^{+4}_{free} = \frac{1}{2}
    q^{+a} [\mathcal{D}^{++},\hat{\Lambda}^{\alpha\beta\dot{\alpha}\dot{\beta}}]\partial_{\alpha\dot{\alpha}}\partial_{\beta\dot{\beta}} q^+_a - \frac12 \Omega_{(4)}
    (q^{+a} {\cal D}^{++} q^+_a)\nn
&&+\,\frac12 (\partial_{\alpha\dot{\alpha}} \lambda^{(\alpha\beta)(\dot\alpha\dot\beta)M})\big[ (\partial_{\beta\dot\beta}q^{+a})\partial_M {\cal D}^{++}q^+_a
+ (\partial_M q^{+a})\partial_{\beta\dot\beta}{\cal D}^{++}q^+_a \big] \nn
&&+\,\frac12 \Omega^{\alpha\beta\dot\alpha\dot\beta} (\partial_{\alpha\dot\alpha} q^{+a}) \partial_{\beta\dot\beta} {\cal D}^{++}q^+_a \label{var5mod}
\eea
(the variation \eqref{var5} is reproduced by taking off all derivatives from ${\cal D}^{++}q^+_a$ in \eqref{var5mod}, except for the first term).\\

\noindent{\it \underline{Invariant Lagrangian}.} Using the formulas for variations and summing them with undetermined coefficients, one can single out their unique combination
\begin{equation}
    \delta_{\lambda} q^{+a}
    :=
    \frac12\left(\delta_1 + \delta_3 +  \delta_4  \right) q^{+a}\,,
\end{equation}
which can be canceled  by the gauge transformation of $\mathcal{N}=2$ invariant analytic differential operator $\hat{\mathcal{H}}^{++(\alpha\beta)(\dot{\alpha}\dot{\beta})}$,
constructed out of the analytic gauge prepotentials:
\begin{equation}
    \begin{split}
    \delta_\lambda \mathcal{L}^{+4}_{free} =& \frac12\left(\delta_1 + \delta_3 +  \delta_4 \right) \mathcal{L}^{+4}_{free}
    =
      \frac12  q^{+a}  [\mathcal{D}^{++}, \hat{\Lambda}^{\alpha\beta\dot{\alpha}\dot{\beta}}] \partial_{\alpha\dot{\alpha}} \partial_{\beta\dot{\beta}} q^+_a
    \\=&
      \frac12\,  q^{+a}  \delta \hat{\mathcal{H}}^{\alpha\beta\dot{\alpha}\dot{\beta}} \partial_{\alpha\dot{\alpha}} \partial_{\beta\dot{\beta}} q^+_a\,.
        \end{split}
\end{equation}
The factor $\frac12$ was introduced for ensuring that $\delta_\lambda q^{+a}$ is reduced to \eqref{global4} in the rigid limit.

Thus the Lagrangian describing the coupling of spin 4 gauge supermultiplet to the hypermultiplet, in the leading order in the former, reads:
\begin{equation}\label{spin4c}
        \mathcal{L}^{+4(s=4)}_{gauge} = -   \frac12\, q^{+a}  \left(\mathcal{D}^{++} + \hat{\mathcal{H}}^{++}_{(4)} \right) q^+_a\,.
\end{equation}
The Lagrangian $\mathcal{L}^{+4}_{gauge}$ is gauge invariant (in the leading order) and completely  $\mathcal{N}=2$ supersymmetric. In contrast to the spin ${\bf s}=3$ case,
the option of adding terms with $\Gamma^{++}_{(4)}$ is absent  in the ${\bf s}=4$ theory.

For further generalizations, it is useful to rewrite $\delta_{\lambda} q^{+a}$ as:
\begin{equation}\label{tr4-final}
    \begin{split}
        \delta_{\lambda} q^{+a}
        = -\frac12\left\{\hat{\Lambda}^{(\alpha\beta)(\dot{\alpha}\dot{\beta})}, \partial_{\alpha\dot{\alpha}} \partial_{\beta\dot{\beta}} \right\} q^{+a}
        -
        \frac12\partial_{\alpha\dot{\alpha}} \partial_{\beta\dot{\beta}}
        \Omega^{(\alpha\beta)(\dot{\alpha}\dot{\beta})}   q^{+a}\,,
    \end{split}
\end{equation}
or, in a more explicit form,
\bea
&&\delta_{\lambda} q^{+a} = -\Big[ \lambda^{(\alpha\beta)(\dot\alpha\dot\beta) M}\partial_{\alpha\dot\alpha}\partial_{\beta\dot\beta} +
(\partial_{\alpha\dot\alpha}\lambda^{(\alpha\beta)(\dot\alpha\dot\beta) M})\partial_{\beta\dot\beta}
+  \frac12(\partial_{\beta\dot\beta}\partial_{\alpha\dot\alpha}\lambda^{(\alpha\beta)(\dot\alpha\dot\beta) M})\Big]\partial_M q^{+a} \nn
&&-\, \Big[(\partial_{\alpha\dot\alpha}\Omega^{(\alpha\beta)(\dot\alpha\dot\beta)})\partial_{\beta\dot\beta} +
\frac12\Omega^{(\alpha\beta)(\dot\alpha\dot\beta)}\partial_{\beta\dot\beta}\partial_{\alpha\dot\alpha} +\Omega_{(4)} \Big] q^{+a}\,.
\eea

We stress once more that the Lagrangian \eqref{spin4c}, like its  spin ${\bf 3}$ counterpart \eqref{spin3-gauge}, is invariant under gauge transformation only in
the leading order in the gauge prepotentials. As distinct from the spin ${\bf 2}$ case, no clear way is seen how to modify the linearized transformation laws \eqref{Gauge_s} so as
to restore the full spin ${\bf s}=4$ gauge invariance of the hypermultiplet coupling (if exists).

\section{Generalization to arbitrary ${\cal N}=2$ spins}
 \label{sec:hyperGen}

\subsection{General odd spins}
\label{sec:41}
In this section, we will assume that ${\bf s}$ is odd integer.
As before, we start with {\it \underline{rigid transformations}}:
\begin{equation}
    \begin{split}
    \delta^{(s)}_{rig} q^{+a} =&- \hat{\Lambda}_{rig}^{\alpha(s-2)\dot{\alpha}(s-2)}\partial^{(s-2)}_{\alpha(s-2)\dot{\alpha}(s-2)} J q^{+a}
    \\=& -i (\tau^3)^a_{\;\;b}\hat{\Lambda}_{rig}^{\alpha(s-2)\dot{\alpha}(s-2)}\partial^{(s-2)}_{\alpha(s-2)\dot{\alpha}(s-2)} q^{+b}\,,
    \end{split}
\end{equation}
where
\begin{equation}\label{Lambda-rigid}
    \hat{\Lambda}_{rig}^{\alpha(s-2)\dot{\alpha}(s-2)} := \Lambda^{\alpha(s-2)\dot{\alpha}(s-2)M} \partial_M\,,
\end{equation}
and
\begin{subequations}\label{LambdA}
\begin{equation}
    \Lambda^{\alpha(s-2)\dot{\alpha}(s-2)\alpha\dot\alpha} =
    \lambda^{\alpha(s-1)\dot{\alpha}(s-1)} - 2i \lambda^{\alpha(s-1)(\dot{\alpha}(s-2)-} \bar{\theta}^{+\dot{\alpha})}
    -
    2i
    \theta^{+(\alpha} \lambda^{\alpha(s-2)) \dot{\alpha}(s-1)-}\,,
\end{equation}
\begin{equation}
    \Lambda^{\alpha(s-2)\dot{\alpha}(s-2)5} =
    \lambda^{\alpha(s-2)\dot{\alpha}(s-2)5}
    +
    2i \lambda^{(\alpha(s-2)\beta)\dot{\alpha}(s-2)-} \theta^+_{{\beta}} + 2i \lambda^{\alpha(s-2)(\dot{\alpha}(s-2)\dot\beta) -} \bar\theta^+_{{\dot\beta}}\,,
\end{equation}
\begin{equation}
    \Lambda^{\alpha(s-1)\dot{\alpha}(s-2)+}
    =
    \lambda^{\alpha(s-1)\dot{\alpha}(s-2)+}\,,
    \;\;\;\;\;
    \lambda^{\alpha(s-1)\dot{\alpha}(s-2)\pm}
    =
    \lambda^{\alpha(s-1)\dot{\alpha}(s-2)i} u_i^\pm
\end{equation}
\begin{equation}
    \Lambda^{\alpha(s-2) \dot{\alpha}(s-1)+}=
    \lambda^{\alpha(s-2) \dot{\alpha}(s-1)+}\,,
    \;\;\;\;\;
    \lambda^{\alpha(s-2) \dot{\alpha}(s-1)\pm}=
    \lambda^{\alpha(s-2) \dot{\alpha}(s-1)i} u_i^\pm\,.
\end{equation}
\end{subequations}
Here all the parameters $\lambda^{\alpha(s-1)\dot{\alpha}(s-1)}\,, \lambda^{\alpha(s-2)\dot{\alpha}(s-2)5}\,,\, \lambda^{\alpha(s-1)\dot{\alpha}(s-2)i},
\,\lambda^{\alpha(s-2) \dot{\alpha}(s-1)i}$ in \break $\hat{\Lambda}_{rig}^{\alpha(s-2)\dot{\alpha}(s-2)}$ are coordinate-independent. We have $s^2 + (s-1)^2$ bosonic parameters
and $2s(s-1)$ fermionic ones. There is valid the property:
\begin{equation}
    [\mathcal{D}^{++},  \hat{\Lambda}_{rig}^{\alpha(s-2)\dot{\alpha}(s-2)}] \partial_{\alpha(s-2)\dot{\alpha}(s-2)}^{s-2} = 0\,.
\end{equation}
The variation of the free hypermultiplet Lagrangian (modulo total derivatives) is zero due to the symmetry of $(\tau^3)_{ab} $:
\begin{equation}
    \begin{split}
    \delta^{(s)}_{rig} \mathcal{L}^{+4}_{free}
    =&
    -i\frac{1}{2} (\tau^3)_{ab} \hat{\Lambda}_{gl}^{\alpha(s-2)\dot{\alpha}(s-2)}\partial^{(s-2)}_{\alpha(s-2)\dot{\alpha}(s-2)} q^{+a} \mathcal{D}^{++} q^{+b}
    \\&+i
    \frac{1}{2} (\tau^3)_{ab}  q^{+a} \mathcal{D}^{++} \hat{\Lambda}_{gl}^{\alpha(s-2)\dot{\alpha}(s-2)}\partial^{(s-2)}_{\alpha(s-2)\dot{\alpha}(s-2)} q^{+b} = 0.
        \end{split}
\end{equation}
To check this, one needs to integrate $s-1$ times by parts in the second term.\\

\noindent{\it \underline{Gauging}}. Like in the spin ${\bf 3}$ case, one
could start with the most general combination of gauge
transformation and then follow the strategy of section
\ref{3.2}. But due to the strong constraints
\eqref{constraints} on the form of interaction, it will be
sufficient to guess, from the very beginning, the gauge
transformations of hypermultiplet which lead to the gauge-invariant
coupling with higher spins. The interactions obtained in this way
will be unique. Fortunately, the sought gauge transformations  prove
to be a direct generalization of those we met while considered the
spin ${\bf 3}$ and spin ${\bf 4}$ examples.

In the generic case of the odd spin supermultiplet, the proper generalization of the spin ${\bf 3}$ gauge
transformations  \eqref{tranf-spin3-final}
is as follows

\begin{equation}\label{sspin1}
    \delta^{(s)}_{\lambda,1} q^{+a} = -i\frac{1}{2}(\tau^3)^a_{\;\;b} \left\{\partial^{(s-2)}_{\alpha(s-2)\dot{\alpha}(s-2)}, \hat{\Lambda}^{\alpha(s-2)\dot{\alpha}(s-2)} \right\} q^{+b}\,.
\end{equation}

The variation of the hypermultiplet Lagrangian can be calculated in full analogy with \eqref{var1}:
\begin{equation}\label{k-spin-11}
    \begin{split}
        \delta_{\lambda,1}^{(s)} \mathcal{L}^{+4}_{free} =& -i  \frac{1}{4} (\tau^3)_{ab} \left\{\partial^{(s-2)}_{\alpha(s-2)\dot{\alpha}(s-2)}, \Omega^{\alpha(s-2)\dot{\alpha}(s-2)} \right\} q^{+a} \mathcal{D}^{++} q^{+b}
        \\&+i \frac{1}{2}  (\tau^3)_{ab}  q^{+a}  \left[\mathcal{D}^{++},\hat{\Lambda}^{\alpha(s-2)\dot{\alpha}(s-2)} \right]  \partial^{(s-2)}_{\alpha(s-2)\dot{\alpha}(s-2)}  q^{+b}
        \\&-i
        \frac{1}{4} (\tau^3)_{ab}
        \left(\mathcal{D}^{++} \Omega^{\alpha(s-2)\dot{\alpha}(s-2)}\right)
        \partial^{(s-2)}_{\alpha(s-2)\dot{\alpha}(s-2)}  q^{+a} q^{+b}.
    \end{split}
\end{equation}

Now we are led to find a transformation that would cancel the first term in \eqref{k-spin-11}.
It is a direct generalization of the second term in \eqref{tranf-spin3-final}:
\begin{equation}\label{var222222}
    \delta^{(s)}_{\lambda,2} q^{+a} = i \frac{1}{2} (\tau^3)^a_{\;b} \partial^{(s-2)}_{\alpha(s-2)\dot{\alpha}(s-2)} \Omega^{\alpha(s-2)\dot{\alpha}(s-2)}  q^{+b} \,.
\end{equation}
The relevant variation of the free hypermultiplet action, in full analogy with the spin ${\bf 3}$ case, is
\begin{equation}\label{kspin22}
    \begin{split}
        \delta_{\lambda,2}^{(s)} \mathcal{L}^{+4}_{free} =&\,
        i\frac{1}{4} (\tau^3)_{ab} \{\partial^{(s-2)}_{\alpha(s-2)\dot{\alpha}(s-2)},  \Omega^{\alpha(s-2)\dot{\alpha}(s-2)} \}  q^{+a} \mathcal{D}^{++} q^{+b}
        \\&+i
        \frac{1}{4} (\tau^3)_{ab}
        \left(\mathcal{D}^{++} \Omega^{\alpha(s-2)\dot{\alpha}(s-2)}\right)
        \partial^{(s-2)}_{\alpha(s-2)\dot{\alpha}(s-2)}  q^{+a} q^{+b}\,.
    \end{split}
\end{equation}

Using \eqref{kspin22}, one can cancel the  first and last terms in \eqref{k-spin-11}. Collecting all terms, we obtain for the full variation :
\begin{equation}
    \begin{split}
        \left( \delta_{\lambda,1}^{(s)} + \delta_{\lambda,2}^{(s)} \right) \mathcal{L}^{+4}_{free}
        =\,&
        i \frac{1}{2}  (\tau^3)_{ab}  q^{+a}  \left[\mathcal{D}^{++},\hat{\Lambda}^{\alpha(s-2)\dot{\alpha}(s-2)} \right]  \partial^{(s-2)}_{\alpha(s-2)\dot{\alpha}(s-2)}  q^{+b}
        \\=\,&
        i \frac{1}{2}  (\tau^3)_{ab}  q^{+a}  \delta\hat{\mathcal{H}}^{++}_{(s)}  q^{+b}\,.
    \end{split}
\end{equation}

Using the transformation law \eqref{transf}, one can cancel these terms by passing to the gauge superfield-modified Lagrangian:
\begin{equation}\label{odd}
    \begin{split}
        \mathcal{L}^{+4(odd\;s)}_{gauge} & =- \frac{1}{2} q^{+a} \left(\mathcal{D}^{++} + \hat{\mathcal{H}}^{++}_{(s)} J  \right) q^+_a\,.
    \end{split}
\end{equation}

The relevant action is a direct generalization of the corresponding action for ${\bf s}=3$ defined by the Lagrangian \eqref{spin3-gauge}. The Lagrangian\eqref{odd}
is also invariant only up to the leading order in gauge prepotentials.

Let us now turn to the $\xi$-transformations \eqref{xi}. A direct generalization of the latter is
\begin{equation}\label{xi-s}
    \delta^{(s)}_{\xi} q^{+a} = - \xi \left( \partial^{s-2}_{\alpha(s-2)\dot{\alpha}(s-2)} \Omega^{\alpha(s-2)\dot{\alpha}(s-2)} \right) J q^{+a}\,.
\end{equation}
The relevant variation of the free Lagrangian has the form:
\begin{equation}
    \delta^{(s)}_{\xi} \mathcal{L}^{+4}_{free} = i \frac{1}{2}\xi\; \left( \mathcal{D}^{++} \partial^{s-2}_{\alpha(s-2)\dot{\alpha}(s-2)} \Omega^{\alpha(s-2)\dot{\alpha}(s-2)} \right)   (\tau^3)_{ab}  q^{+a} q^{+b}
\end{equation}
and it can be easily canceled by adding $\Gamma^{++}_{(s)}$ \eqref{Gamma-main}, so the most general Lagrangian reads:
\begin{equation}\label{odd-final}
    \begin{split}
        \mathcal{L}^{+4(odd\;s)}_{gauge}  = -\frac{1}{2} q^{+a} \left(\mathcal{D}^{++} + \hat{\mathcal{H}}^{++}_{(s)} J
        +
        \xi \Gamma^{++}_{(s)} J \right) q^+_a\,.
    \end{split}
\end{equation}
Here $\xi$ is an arbitrary real parameter.

To summarize, for an arbitrary odd ${\cal N}=2$ spin ${\bf s}$ the gauge transformations of the hypermultiplet  are given by $\delta_\lambda^{(s)} q^{_a} =\delta_{\lambda, 1}^{(s)} q^{+a}
+ \delta_{\lambda, 2}^{(s)} q^{+a}$ and $\delta^{(s)}_{\xi}$, with the variations being defined in \eqref{sspin1}, \eqref{var222222} and \eqref{xi-s}. The action which is  gauge invariant up
to the first order in gauge superfields is given by eq. \eqref{odd-final}. The hypermultiplet can be massless ($\partial_5 q^{+ a} =0\,, \; J q^{+ a} \neq 0$) or massive
($\partial_5 q^{+ a} \neq 0\,, \; J q^{+ a} \neq 0\,, \; \partial_5 q^{+ a} = m J q^{+ a} $).

\subsection{General even spins}
In this section, we will generalize the spin ${\bf 4}$ transformation laws \eqref{tr4-final} to an arbitrary integer even spin ${\bf s}$.

Let us start with the definition of {\it \underline{rigid symmetry}} of the free hypermultplet for even higher spins:
\begin{equation}\label{global-even}
    \delta_{rig}^{(s)} q^{+a} =  -\hat{\Lambda}_{rig}^{\alpha(s-2)\dot{\alpha}(s-2)} \partial^{s-2}_{\alpha(s-2)\dot{\alpha}(s-2)} q^{+a}\,.
\end{equation}
The group-parameter structure of
\begin{equation}
    \hat{\Lambda}_{rig}^{\alpha(s-2)\dot{\alpha}(s-2)} = \Lambda^{\alpha(s-2)\dot{\alpha}(s-2)M} \partial_M
\end{equation}
coincides with that  already given in \eqref{LambdA}, the crucial difference being the absence of the generator $J$ (and, respectively, of the matrix $(\tau^3)^a_b$)
in the transformation law \eqref{global-even}.
It is a direct generalization of the rigid symmetry \eqref{global4} pertinent to the spin ${\bf s}=4$.
Here we face $s^2 + (s-1)^2$ constant bosonic parameters and $2s(s-1)$ constant fermionic parameters. One can check that this transformation, in the complete analogy
with \eqref{global-spin4}, indeed provides a symmetry of the free hypermultiplet action.  Now we will gauge this symmetry, using the even higher-spin  ${\bf s}$ gauge supermultiplet.\\

\noindent{\it \underline{Gauging}}. Like in the previous cases, one can define many local generalizations of the rigid transformation \eqref{global-even},
with the proper analytic superfield parameters and derivatives thereof. However,  as a consequence of the strong restrictions \eqref{constraints} on the possible structure
of the interaction, it is enough to explicitly guess some kind of implementation of the gauge group, such that it meets the restrictions just mentioned.
The interaction constructed in this way will be most general. The option leading to the desired result is a direct generalization of the spin 4 gauge transformations \eqref{tr4-final}:
\begin{equation}\label{trs-final}
    \begin{split}
        \delta^{(s)}_{\lambda} q^{+a}
        &= -\frac{1}{2}\left\{\hat{\Lambda}^{\alpha(s-2)\dot{\alpha}(s-2)}, \partial^{s-2}_{\alpha(s-2)\dot{\alpha}(s-2)}\right\} q^{+a}
        - \frac{1}{2}
        \partial^{s-2}_{\alpha(s-2)\dot{\alpha}(s-2)} \Omega^{\alpha(s-2)\dot{\alpha}(s-2)}   q^{+a}\,.
    \end{split}
\end{equation}

To explicitly check that the transformation \eqref{trs-final} is the needed one, we divide it into two terms. The first part of the variation,
\begin{equation}
    \delta^{(s)}_{\lambda,1} q^{+a}
    = -\frac{1}{2}\left\{\hat{\Lambda}^{\alpha(s-2)\dot{\alpha}(s-2)}, \partial^{s-2}_{\alpha(s-2)\dot{\alpha}(s-2)}\right\} q^{+a}\,,
\end{equation}
transforms the free hypermultiplet Lagrangian (up to a total derivative) as
\begin{multline}
    \delta_{\lambda, 1}^{(s)} \mathcal{L}^{+4}_{free}
    =
    -\frac{1}{4}
    \left\{\Omega^{\alpha(s-2)\dot{\alpha}(s-2)}, \partial^{s-2}_{\alpha(s-2)\dot{\alpha}(s-2)}\right\} q^{+a} \mathcal{D}^{++} q^+_a
    \\+\frac{1}{2}
    q^{+a} [\mathcal{D}^{++} , \hat{\Lambda}^{\alpha(s-2)\dot{\alpha}(s-2)}] \partial^{s-2}_{\alpha(s-2)\dot{\alpha}(s-2)} q^+_a
    +
    \frac{1}{4}\left(\mathcal{D}^{++}  \Omega^{\alpha(s-2)\dot{\alpha}(s-2)}\right)
    q^{+a}  \partial^{s-2}_{\alpha(s-2)\dot{\alpha}(s-2)} q^+_a   \,.
\end{multline}

The second part of the transformation \eqref{trs-final},
\begin{equation}
    \delta^{(s)}_{\lambda,2} q^{+a}
    =
    - \frac{1}{2}
    \partial^{s-2}_{\alpha(s-2)\dot{\alpha}(s-2)} \Omega^{\alpha(s-2)\dot{\alpha}(s-2)}   q^{+a}\,,
\end{equation}
gives rise to the following transformation of the free hypermultiplet Lagrangian:
\begin{equation}
    \begin{split}
    \delta_{\lambda, 2}^{(s)} \mathcal{L}^{+4}_{free} =\,&\frac{1}{4}
    \left\{\Omega^{\alpha(s-2)\dot{\alpha}(s-2)}, \partial^{s-2}_{\alpha(s-2)\dot{\alpha}(s-2)}\right\} q^{+a} \mathcal{D}^{++} q^+_a
    \\&-
    \frac{1}{4}\left(\mathcal{D}^{++}  \Omega^{\alpha(s-2)\dot{\alpha}(s-2)}\right)
    q^{+a}  \partial^{s-2}_{\alpha(s-2)\dot{\alpha}(s-2)} q^+_a   \,.
        \end{split}
\end{equation}
So the full variation of the Lagrangian under \eqref{trs-final} allows one to determine the possible interaction terms:
\begin{equation}
    \delta_{\lambda}^{(s)} \mathcal{L}^{+4}_{free} = \frac{1}{2} q^{+a} [\mathcal{D}^{++} , \hat{\Lambda}^{\alpha(s-2)\dot{\alpha}(s-2)}] \partial^{s-2}_{\alpha(s-2)\dot{\alpha}(s-2)} q^+_a
    =
    \frac{1}{2} q^{+a} \delta \hat{\mathcal{H}}^{++}_{(s)} q^+_a\,.
\end{equation}
As a consequence, the gauge invariant Lagrangian has the form:
\begin{equation}\label{spinSc}
    \mathcal{L}^{+4(even\;s)}_{gauge}
    =  - \frac{1}{2}    q^{+a}  \left(\mathcal{D}^{++} +\hat{\mathcal{H}}^{++}_{(s)}  \right) q^+_a\,.
\end{equation}
Like in all previous cases, it is the covariantization of the free hypermultiplet action under \eqref{trs-final}
through extending the harmonic derivative $\mathcal{D}^{++}$ by the differential operator $\hat{\mathcal{H}}^{++}_{(s)}$ \eqref{operHS}.
The highest derivative term in this operator have the degree $(s-1)$. Once again, the action corresponding to \eqref{spinSc} is invariant only to
the first-order in gauge superfields. For the time being, we do not know how to achieve the complete invariance of such an action.

\section{Summary and outlook}
\label{sec:summary}
Here we briefly summarize and discuss the results obtained.\\

\noindent\textbf{1.} First of all, we have identified the infinite dimensional {\it rigid symmetry of hypermultiplet} realized by higher-derivative transformations.
These can be written in the universal form, at once for the odd and even spins:
\begin{equation}\label{global-sym}
    \delta^{(s)}_{rig} q^{+a} = -\hat{\Lambda}_{rig}^{\alpha(s-2)\dot{\alpha}(s-2)}\partial^{s-2}_{\alpha(s-2)\dot{\alpha}(s-2)} (J)^{P(s)} q^{+a}\,,
    \quad P(s) = \frac{1 + (-1)^{s+1}}{2}\,,
\end{equation}
\begin{equation}
    [\mathcal{D}^{++}, \hat{\Lambda}_{rig}^{\alpha(s-2)\dot{\alpha}(s-2)}] \partial^{s-2}_{\alpha(s-2)\dot{\alpha}(s-2)} = 0\,.
\end{equation}
The parameters $\hat{\Lambda}_{rig}^{\alpha(s-2)\dot{\alpha}(s-2)}$ were defined in \eqref{Lambda-rigid} and \eqref{LambdA}. The transformation $\delta^{(s)}_{rig}$ contains $(s-1)$ pure vector derivatives $\partial^{s-1}_{\alpha(s-1)\dot{\alpha}(s-1)}$   and  $(s-2)$ vector derivatives
times the spinor derivative $\partial^-_{\hat{\mu}}$. It involves $s^2 + (s-1)^2$ bosonic parameters and $2s(s-1)$ fermionic parameters, total of $(2s-1)^2$ parameters.

It is worth noting that the group of rigid symmetries of the free hypermultiplet is much wider than \eqref{global-sym}.
For any spin there are also transformations of the form:
\begin{equation}\label{general}
    \delta^{(s)}_{general(1)} \; q^{+a} = -\hat{\Lambda}_{\;\;b}^{a\;\alpha(s-2)\dot{\alpha}(s-2)}\partial^{s-2}_{\alpha(s-2)\dot{\alpha}(s-2)} q^{+b}\,,
\end{equation}
involving the matrix parameter $\hat{\Lambda}_{\;\;b}^{a\;\alpha(s-2)\dot{\alpha}(s-2)}$ which is determined by the same formulas \eqref{Lambda-rigid} and \eqref{LambdA},
with extra doublet ${\rm SU}(2)_{PG}$ indices $a$ and $b$. This parameter is symmetric(antisymmetric) depending on whether $s$ is odd(even):
\begin{equation}\label{Lambda-general}
    \hat{\Lambda}_{ab}^{\alpha(s-2)\dot{\alpha}(s-2)}
    =
    (-1)^{s+1}\hat{\Lambda}_{ba}^{\alpha(s-2)\dot{\alpha}(s-2)}\,.
\end{equation}

To be convinced of this, consider a variation of the free hypermultiplet action of the hypermultiplet with respect to \eqref{general}:
\begin{equation}
    \begin{split}
    \delta^{(s)}_{general(1)} \mathcal{L}^{+4}_{free}
    =
    &-\frac{1}{2} \left(\hat{\Lambda}_{ab}^{\;\alpha(s-2)\dot{\alpha}(s-2)}\partial^{s-2}_{\alpha(s-2)\dot{\alpha}(s-2)} q^{+b}\right)
    \mathcal{D}^{++} q^{+a}
    \\&+
    \frac{1}{2}  q^{+b}
    \mathcal{D}^{++} \left(\hat{\Lambda}_{ba}^{\;\alpha(s-2)\dot{\alpha}(s-2)}\partial^{s-2}_{\alpha(s-2)\dot{\alpha}(s-2)} q^{+a}\right)\,.
    \end{split}
\end{equation}
After integration by parts, one can rewrite this variation as:
\begin{equation}
    \delta^{(s)}_{general(1)} \mathcal{L}^{+4}_{free}
    =
    -\frac{1}{2} \left(\left[\hat{\Lambda}_{ab}^{\;\alpha(s-2)\dot{\alpha}(s-2)} + (-1)^s \hat{\Lambda}_{ba}^{\;\alpha(s-2)\dot{\alpha}(s-2)} \right] \partial^{s-2}_{\alpha(s-2)\dot{\alpha}(s-2)} q^{+b}\right)
    \mathcal{D}^{++} q^{+a}\,.
\end{equation}
Then the requirement of invariance of the action amounts to the condition \eqref{Lambda-general}.

For even spins ${\bf s} \geq 2$, this does not provide new possibilities as compared to
\eqref{global-sym}, while for odd spins ${\bf s} \geq 3$ new rigid
symmetries come out. These extra rigid symmetries generalize to the case of the
hypermultiplet the symmetries suggested in ref. \cite{Berends:1985xx}. Perhaps, it would be of interest to explore their
gauging. Leaving aside the detailed treatment of this interesting problem, note that
such a gauging would imply the introduction of new gauge superfield prepotentials with a nontrivial index structure.
 This should also lead to a change in the structure of the action of
higher spins and to the appearance of new indices for the component
fields. So gauging of such symmetries could yield some ``non-abelian higher spin theory''.
To the best of our knowledge, even for the bosonic higher spins, the interactions of this kind had never been seriously explored.
The only relevant remark is contained in ref. \cite{Berends:1985xx}, where it is claimed that the gauging of
such symmetries ``run into difficulties''. 

A similar (though different) kind of ``non-abelian'' higher spin invariances of the free hypermultiplet action \eqref{hyper}, also existing for odd spins ${\bf s} = 1, 3,\ldots $,
and generalizing those of ref. \cite{Berends:1985xx}, is provided by the transformations
\be
\delta^{(s)}_{general(2)} \; q^{+}_a = {\Lambda}_{(ab)}^{\alpha(s-1)\dot{\alpha}(s-1)}\partial^{s-1}_{\alpha(s-1)\dot{\alpha}(s-1)} q^{+b}\,,\label{global2}
\end{equation}
where the parameters are $c$-numbers and for ${\bf s}=1$ yield just rigid  ${\rm SU}(2)_{PG}$ transformations. So gauging of such symmetries should result
in some higher-spin generalizations of Yang-Mills theory associated with the group ${\rm SU}(2)_{PG}$. So much for these new opportunities  which
we hope to study elsewhere.\\

\noindent\textbf{2.} As the next crucial step after defining the global symmetries \eqref{global-sym}, we {\it gauged } them and explicitly presented the gauge
transformations of $4D$, $\mathcal{N}=2$
analytic hypermultiplet superfield. They constitute a family of transformations which differ by the highest degree $s$ of the involved derivatives. These transformations
are consistent with the analyticity and include the same differential operators  as the gauge transformations of the higher-superspin analytic prepotentials \eqref{transf}.
These transformations have the universal form:
\begin{equation}\label{transformations}
    \begin{split}
    \delta^{(s)}_{\lambda} q^{+a}
    =& - \frac{1}{2}\left\{\hat{\Lambda}^{\alpha(s-2)\dot{\alpha}(s-2)}, \partial^{s-2}_{\alpha(s-2)\dot{\alpha}(s-2)}\right\} (J)^{P(s)} q^{+a}
    \\&-\frac{1}{2}
    \partial^{s-2}_{\alpha(s-2)\dot{\alpha}(s-2)} \Omega^{\alpha(s-2)\dot{\alpha}(s-2)} (J)^{P(s)}  q^{+a}\,.
        \end{split}
\end{equation}
When the parameters in  \eqref{transformations} are independent of the coordinates, these transformations reproduce the global symmetry group \eqref{global-sym}
of the free hypermultiplet action \eqref{hyper}.\\

\noindent\textbf{3.} As the final result, in section \ref{sec:hyper} we presented $\mathcal{N}=2$ {\it gauge invariant cubic couplings} $\mathbf{\frac{1}{2} - \frac{1}{2} - s}$
of the hypermultiplet to higher-spin $\mathbf{s}$  $\mathcal{N}=2$ gauge supermultiplets, which we described in section \ref{sec:hs}. These couplings
can be brought in the universal form for all spins:
\begin{equation}\label{action-final}
        S^{(s)}_{gauge}
        =
       - \frac{1}{2} \int d\zeta^{(-4)}\;
        q^{+a} \left(\mathcal{D}^{++} + \hat{\mathcal{H}}^{++}_{(s)}  (J)^{P(s)} + \xi \Gamma^{++}_{(s)} (J)^{P(s)} \right) q^+_a\,.
\end{equation}
The actions \eqref{action-final} are gauge-invariant under \eqref{Gauge_s} and \eqref{transformations}
only in the leading order in the gauge superfields, except the cases of $\mathcal{N}=2$ Maxwell supermultiplet  ($\mathbf{s}=1$) and $\mathcal{N}=2$
supergravity Einstein multiplet ($\mathbf{s}=2$). Note that for even ${\bf s}$ the $\xi$ term in \eqref{action-final} disappears. For odd spins the $\xi$ term
can be included and one is led to add, to the transformations \eqref{transformations}, some extra gauge transformation:
\begin{equation}\label{xi-s-f}
    \delta^{(s)}_{\xi} q^{+a} = -   P(s)\xi \left( \partial^{s-2}_{\alpha(s-2)\dot{\alpha}(s-2)} \Omega^{\alpha(s-2)\dot{\alpha}(s-2)} \right)  J q^{+a}\,.
\end{equation}

Thus, the interactions for even and odd spins differ by the presence, in the odd spin case,  of an additional
interaction with the dimensionless coupling constant $\xi$. At the superfield off-shell level, this second interaction
has the form clearly distinct from the basic interaction. The most important difference between these two interactions is that the $\xi$-terms contain no derivatives
on the hypermultiplet superfield. The interplay of these two terms at the on-shell component level will be discussed
elsewhere. Note that after passing to the gauge superfields of the canonical dimension, the constants $\kappa_{s}$ defined in \eqref{ActionsGen}  appear in
\eqref{action-final} in front of the interaction terms, and so the interactions with hypermultiplets vanish in the limit $\kappa_{s} \to 0\,$. \\


\noindent\textbf{4.} The presence of the ${\rm U}(1)_{PG}$ generator  $J $ of the internal ${\rm SU}(2)_{PG}$ symmetry
in the transformations \eqref{transformations} and couplings \eqref{action-final} results in an essential difference in the treatment of the odd and even spin ${\bf s}$ cases.
For odd  ${\bf s}$ the relevant higher-spin transformations
of the hypermultiplet and its couplings to the gauge fields exist only provided ${\rm SU}(2)_{PG}$ is broken, with $\partial_5 q^{+a} = 0\,, \;J q^{+a} \neq 0$ in the
case of massless hypermultiplet and $\partial_5 q^{+a} \neq 0\,, \;J q^{+a} \neq 0\,, \;\partial_5 q^{+a} \sim m J $ in the case of {\it massive} hypermultiplet.
The unbroken  ${\rm SU}(2)_{PG}$ symmetry  implies $\partial_5 q^{+a} = J q^{+a} =0$, so the gauge transformations of the hypermultiplet and the relevant gauge invariant
couplings in this case can be defined only for the even spins ${\bf s}$, the hypermultiplet being massless.

It is the appropriate place here to discuss how our consideration can be extended to the case of few hypermultiplets.
The free Lagrangian of $n$ hypermultiplets can be written in the manifestly ${\rm USp}(2n)$ invariant form as
\bea
{\cal L}^{+4}_{free, n} = \frac12 q^{+ A}{\cal D}^{++}q^+_A\,, \quad \widetilde{q^+_A} = \Omega^{AB} q^+_B\,, \;\; A = 1, 2, \ldots , 2n\,,\label{GenqA}
\eea
where $\Omega^{AB} = -\Omega^{BA}$ is ${\rm USp}(2n)$ invariant constant $2n \times 2n$ symplectic metric. Up to a numerical factor, this Lagrangian can be rewritten in an equivalent complex form as
\bea
{\cal L}^{+4}_{free, n} \sim \tilde{q}^{+a}{\cal D}^{++}q^+_a -{\cal D}^{++} \tilde{q}^{+a} q^+_a\,, \quad  a= 1, 2, \ldots , n\,, \;\;q^+_A = (q^+_a, - \tilde{q}^{+ a})\,.\label{Genqa}
\eea
In this form it is manifestly invariant under the group ${\rm U}(n)= {\rm SU}(n)\times {\rm U}(1) \subset {\rm USp}(2n)$, with respect to which $q^+_a$ and $\tilde{q}^{+ a}$
transform in the fundamental and
co-fundamental representations, while the transformations completing ${\rm U}(n)$ to ${\rm USp}(2n)$ are realized as $\delta\tilde{q}^{+a} = C^{ab} q^+_b$ (and c.c.),
with $C^{ab} = C^{ba}\,$.  In this general case it is natural to identify $J$ with the common
phase ${\rm U}(1) \subset {\rm U}(n)\,,$
\bea
J q^+_a  = i q^+_a\,, \quad J \tilde{q}^{+ a} =  -i \tilde{q}^{+ a}\,. \label{Ident5Gen}
\eea
It breaks ${\rm USp}(2n)$ down to ${\rm U}(n)\subset {\rm USp}(2n) $\footnote{This is maximally symmetric choice. In principle, one could identify $J$ with a combination
${\rm U}(1)_{Cart}$ of the Cartan generators of ${\rm SU}(n)$, thus breaking   ${\rm USp}(2n)$
to ${\rm U}(1)_{Cart} \times {\rm U}(1) \subset {\rm U}(n)$.}.  So in this case the residual group acting on the bosonic fields $f_a^i$ is ${\rm U}(n) \times {\rm SU(2)}_{aut}\,$.
All the higher-spin gauge transformations
derived earlier for a single hypermultiplet and the relevant cubic couplings to the gauge superfields can be directly transferred to this general case. Note that for generic $n > 1$
there are much more possibilities to choose a pair of mutually commuting internal symmetry generators to be identified with $J$ and, by the Scherk- Schwarz mechanism, with  $\partial_5$ .

\medskip

\noindent\textbf{5.} The results on the cubic $\mathcal{N}=2$ supersymmetric couplings of hypermultiplet to higher spins
obtained in the present paper could be further extended along several directions  (besides those already mentioned in the item ${\bf 1}$) :
\begin{itemize}
    \item
    The natural next step is the construction and investigation of $4D, \mathcal{N}=2$ higher-spin supercurrents of the hypermultiplet;

    \item Of primary interest is also the study of the component structure of the interactions constructed. The bosonic physical fields $f^{Ai}$ of $n$
    hypermultiplets ($4n$ independent fields) are transformed according to the bi-fundamental representation of ${\rm USp}(n)\times {\rm SU}(2)_{aut}$ in the massless case or
    ${\rm U}(n)\times {\rm SU}(2)_{aut}$ in the massive one, and it is  tempting to examine how the bosonic subsector of the ${\cal N}=2$  higher spin gauge group acts
    on these fields;

    \item An interesting task is to explore the relationship of the  $\mathcal{N}=2$ couplings presented here with the known $\mathcal{N}=1$ cubic interactions.
    To accomplish this, it is necessary to reduce our $\mathcal{N}=2$ harmonic superspace to $\mathcal{N}=1$ superfields;

    \item The approach under consideration can be applied to interactions of $\mathcal{N}=2$ higher spin theory with the hypermultiplet defined on ${\rm AdS}$
    and more general superconformally-flat  superbackgrounds. The hypermultiplet action in $5D$ ${\rm AdS}$ harmonic superspace was constructed in \cite{Kuzenko:2007aj, Kuzenko:2007vs}
    and can be directly dimensionally reduced to $4D$. It is still unknown how to extend the harmonic superspace construction of the higher spin theories given in \cite{Buchbinder:2021ite}
    to $4D$ ${\rm AdS}$ framework. The linearized action for $\mathcal{N}=2$ supergravity on the ${\rm AdS}$ background was built in \cite{Butter:2010sc}.
    However, this formulation was never generalized to higher spins.

\end{itemize}

\section*{Acknowledgements}
 Work of I.~B. and E.~I. was supported in part by the Ministry of Education
of Russian Federation, project FEWF-2020-0003. E.~I. thanks Misha Vasiliev for useful correspondence.
The authors are grateful to the anonymous referee for useful and suggestive comments.


\end{document}